# Large-area, ensemble molecular electronics: Motivation and challenges

By Ayelet Vilan[a,*], Dinesh Aswal[b] and David Cahen[a]

a)   Department of Materials and Interfaces, Weizmann Institute of Science, Rehovot, Israel

b)   National Physical Laboratory, New Delhi, India

## Abstract

We review charge transport across molecular monolayers, which is central to molecular electronics (MolEl) using large-area junctions (NmJ). We strive to provide a wide conceptual overview of three main sub-topics. First, a broad introduction places NmJ in perspective to related fields of research, and to single molecule junctions (1mJ), in addition to a brief historical account. As charge transport presents an ultra-sensitive probe for the electronic perfection of interfaces, in the second part ways to form both the monolayer and the contacts are described to construct reliable, defect-free interfaces. The last part is dedicated to understanding and analyses of current – voltage (I-V) traces across molecular junctions. Notwithstanding the original motivation of MolEl, I-V traces are often not very sensitive to molecular details and then provide a poor probe for chemical information. Instead we focus on how to analyse the net electrical performance of molecular junctions, from a functional device perspective. Finally, we shortly point to creation of a built-in electric field as a key to achieve functionality, including non-linear current-voltage characteristics that originate in the molecules or their contacts to the electrodes.

## 1. Introduction

Molecular electronics (MolEl) is commonly viewed as a branch of nanotechnology with molecules serving as ultimate nano-sized building blocks of what is often presented as essentially unlimited functionality due to the supposed ability to synthesize them at will. This review is written to provide a newcomer to the field an account of the fascinating challenges confronting MolEl, which are at the frontier of scientific understanding of chemistry and physics at the nano-scale. The size of these challenges may also explain some of the criticisms on MolEl, as it has not (yet) developed into technologies, or directions towards these, and it has, as yet yielded only limited insights into fundamental molecular processes that can be considered to be unique to the approach. Experimentally, reliable molecular devices require controlling the atomic coordinates of both the molecules and the atoms of the contacting leads (electrodes); the large area junctions that are discussed here, also require repetition of thousands to millions of molecules in parallel, in as perfect a manner as possible. Conceptually, MolEl merges two orthogonal descriptions of electronic states, localized molecular orbitals with free electrons moving in energy bands.

The justification and motivation to study MolEl is described here by way of comparison to related scientific fields (section 2.1). As recently several nice reviews on MolEl have appeared,[1-5] we aim here to view the field from a few different angles. First, we mostly focus on large area junctions, rather than single-molecule junctions. These two branches of MolEl are compared in section 2.2: the differences are deeper than commonly assumed and monolayer-based junctions could be more suitable than single-molecule ones in certain cases. Although study of MolEl is almost half a century old, research before 1990 is often disregarded, possibly because of supposedly poor technical quality of the work. As briefly detailed in section 2.3, there is a surprising similarity between the findings of those pioneering works and contemporary ones, pointing to reproducibility of observations on the one hand, and to some persistent obstacles to MolEl becoming a mature research direction, on the other hand.

Except for the introductory section 2, this review emphasizes primarily two fundamental questions: how to preserve the integrity of a monolayer when it becomes part of a solid interface (sections 3-4) and what is the expected electrical functionality of molecules (sections 5-6). The first question includes the various strategies for forming monolayers (section 3.1), and, specifically we consider the combination of chemical forces that provide long-term stability, monolayer homogeneity and diverse chemical functionality (section 3.2). Monolayers, in contrast to single molecules, benefit from the power of advanced surface analysis tools; these complementary characterizations are an inherent part of the study of monolayers for electronic purposes. For a non-expert reader, a brief account of key characterization tools is provided, divided into those that help to know if we 'got the intended monolayer?' (section 3.3) and 'what are its electronic energy characteristics?' (section 3.4). The section concludes by considering in which sense molecular monolayers differ from inorganic thin films, in terms of possible functionalities (section 3.5). Section 4 continues the construction (of molecular junction) aspects, but now with regard to sandwiching the molecules between two solid contacts. It provides a general classification of the numerous existing variations and their generic pros and cons (4.1). We identify the three critical issues that dictate film (non)uniformity, i.e., diffusion, roughness and chemical interactions. These are then discussed in the context of soft, wet interface preparation as commonly used for large-area molecular junctions (4.2). Lastly we touch upon the fundamental dilemma of large-area junctions, i.e., that defects at a level below the limits of chemical detection methods, can be significant for electrical transport across the junctions (4.3).

In section 5 the review switches gears to look at direct charge transport aspects, and we confront the question of how to describe the electrical functionality of a molecular junction (section 5.1). Such a function-oriented goal, i.e., "what can a molecular device do?" differs from the aim of *probing* chemical details, i.e., wanting to find out "what *is* a molecular junction"? After briefly discussing temperature effects (5.2) we show that tunneling-based molecular $dI/dV - V$ traces are roughly parabolic (5.3). This rough approximation is compared to a specific tunneling model based on the Landauer approach (section 5.4) and unique fingerprints of both analyses are provided. One of the goals of section 5 is to promote the use of scaling bias as an *empirical measure* of the junction's voltage response, rather than seeking an accurate marker of molecular energy levels. The technical extraction of the scaling bias and its comparison to another technique known as 'transition voltage', are covered in section 5.5. Thus, section 5 provides detailed tools to describe charge transport by tunneling across generic molecules. Specific molecular identity is briefly considered in Section 6, through three selected aspects. Within the functional perspective, we ask what causes an asymmetric bias response (6.1). Section 6.2 suggests the answer lies less in the intrinsic molecular energy levels and more in the energy alignment with the contacts and the creation of an intrinsic electric field. For completeness' sake, a brief account of the intensely studied effect of molecular length is provided (6.3).

## 2. Molecular electronics in perspective: Why, which, who?

The study of electrical uses of mono-molecular thick layers (monolayers) as interfaces is part of a general quest towards molecular electronics (MolEl), or the desire to tailor electronic behaviour of a junction starting from basic, smallest building blocks, molecules in this case. This introductory section is meant to place the study of molecular arrays within a broader context of materials for electronics. We shall first compare MolEl to adjacent scientific fields to exemplify the unique features of MolEl (2.1). We then consider the different divisions within MolEl (2.2) and conclude this section by providing a short account of early achievements (2.3).

## 2.1. Why: Benchmarking Molecular Electronics with related scientific fields

The following scientific fields have close relevance to MolEl:

2.1.1. **Mesoscopic physics** focuses on unique quantization effects emerging where the material (potential well) is smaller than some fundamental lengths, such as the elastic mean free path or the de Broglie wavelength of electrons in the leads.[6,7] Roughly speaking, molecules have more localized ("rigid") electrons, but more flexible nuclei than non-molecular solids.[8] Electron localization implies smaller electrostatic screening than in inorganic electronic materials, which helps build-up of steep electrical potential gradients, but may hamper charging events as in, e.g., Coulomb blockade. The motions of the nuclei are slow, compared to electron-crossing rates, and therefore probably do not intervene directly with the transport event. However, nuclei flexibility often implies varying molecular structures in the same nominal junctions, leading to poor reproducibility as well as sensitivity to environment and exact preparation conditions.

2.1.2. **Organic electronics** refers to electronic devices made of conjugated small molecules (dyes) or polymers, normally in the form of thin films (≥ 50 nm thick).[9] In contrast to such systems, the transport distance across molecular junctions is much shorter (1-5 nm) than across organic electronic ones. For this reason, the transport mechanism is generally different between MolEl (fast, electron tunneling) and organic electronics (slow, polaron hopping),[8] with a transition at a certain length.[10] At the same time electrostatic aspects, such as energy alignment, are generally similar for the two fields.[8] The short transport distance allows for a variety of chemical structures that is not limited to conjugated segments and, thus, richer than is the case for organic electronics. A major original goal of MolEl was to have the complete functionality (e.g., a 'diode' or 'switch') within the same molecule,[1,11] different from organic or inorganic thin-film electronics, where often multiple layers of different components serve this purpose.[9,12]

2.1.3. **Electrochemistry** studies charge injection to / from solid electrodes leading to redox reactions occurring at those electrodes. The lack of a solvent as an active component in MolEl is a critical difference with electrochemistry, because solvent polarizability is a vital component for enabling the redox reaction with reasonable activation energies (within the stability window of the molecules).[13,14] Although MolEl shares many concepts with electrochemistry,[15,16] the latter has clear limitations in terms of miniaturization, cooling to low temperatures and/or compatibility with common electrical devices. Still, in MolEl electrochemistry is used to achieve gating of the electron transport across the molecules.[2]

2.1.4. **Electron transfer** (ET) **vs. Electron Transport** (ETp):[14,17,18] the main difference between these concepts as used here, relates to electronic – nuclear coupling: it is weak in electron transport, as in metals and in practical semiconductors, while it is strong and critical for electron transfer, as described by Marcus' ET model for (bio)molecules and electro-chemistry, or in polaronic transport (organic electronics, oxides).[8] Though, a tunneling step is included in the Marcus ET model and the probability of an electron to cross a given molecule (wire) is common to both views,[17,19] an obvious difference is that in ET a rate of *transient* events is measured, while MolEl deals normally with *steady-state* current flow. While a redox process can be detected in MolEl, as spikes in the noise,[20,21] in ET the redox process is detected as a pronounced feature. Another key difference is the nature of the source and final states of the charge. Electron transfer is from one specific electron level (donor) to another (acceptor), occurring only at the occasion that nuclear vibrations bring these two levels close in energy (resonance). In MolEl, however, charges are provided and absorbed by the continuum of states of the electrodes. In the case of tunneling, electron transfer is assumed to be too fast to be affected by nuclear vibrations.[22] In terms of contacts electrochemistry, using adsorbed molecules as the object of study, is intermediate between ET and ETp, because one of the contacts is an electrode with a continuum of states, while the other contact is the electrolyte solution. Such solution will contain redox-active species with defined electron energy levels (broadened by the nuclear vibrations). As electron transfer to the latter is normally rate-limiting, what is measured is an electro-chemical redox event (the above-mentioned pronounced feature; e.g., the "duck" in a cyclic voltammogram), normally absent in ETp. [23]

## 2.2. Which: Large area vs. isolated molecular junctions

This variety of related fields leads to diverse motivations for, and experimental approaches to MolEl. Still, there is one major division in all approaches, that between single molecule junctions[2,24,25] and molecular array-based ones.[1,19,26-29] Molecular arrays are also known as mon-olayers or ensemble junctions, and will be abbreviated as **NmJ**. Historically, measuring charge transport across NmJ pre-dated that of single molecules (**1mJ**), which required the invention of scanning probe microscopy and advanced lithography. Still, NmJ are not lesser predecessors of 1mJ,[30] but rather offer a complementary approach to that of 1mJ and, naturally, the promise of practical products.[31] The following part compares the similarities and differences between these two experimental approaches.

Conceptually, a single molecule is a zero-dimensional object (0D) with dimensions smaller than the electron wavelength, while a monolayer has two dimensions (2D). Nevertheless, in both configurations the separation between the electrodes is normally less than $\sim 5$ nm, i.e., charge transport occurs across a '0 D' also in NmJ. There are exceptions where transport occurs over longer molecular wires[32,33] or thicker films[29,34] and this normally involves hopping transport.[35] Another known exception is monolayer-based *transistors*,[36-39] where charge propagates along the lateral plane of a monolayer, acting as a channel in a field-effect transistor (FET)-like configuration. In a way, that is the ultimate miniaturization of **organic electronics**, and it resembles studies on graphite, graphene and other layered compounds. Ignoring these long-range scenarios, a useful starting point to describe transport across a monolayer (NmJ) is that it is roughly a super-position of numerous, parallel 'single molecule events' (1mJ).[40] This view is not so obvious, because confinement of the electron wave function in all three dimensions is a fundamental aspect for both mesoscopic physics and ET. It is still unclear in how far the electron wavefunction is laterally constrained into a single molecule when these are close-packed within a monolayer and if any generalization is possible.

Possibly the 1mJ vs. NmJ divide is related to differences in motivation. Does MolEl aim at fundamental insight or functional impact? (The obvious answer is "both"[41]). What holds *the* key to functionality, intrinsic or collective molecular properties, interfacial or structural features? Obviously these divisions are at least in part artificial as important insights emerge from combining input from both types of junctions; also, net performance is often a function of a combination of molecule- and contact-related factors.[30,42,43] Still, the two fundamental experimental approaches, 1mJ and NmJ have inherent pros and cons as detailed below.

**2.2.1. Molecules as isolated quantized objects**: the assumption that intrinsic molecular properties control transport originates from the quest for quantized effects. These should be clearest for isolated molecules as in 1mJ. Trapping a single molecule eliminates interactions with neighboring molecules, and is the best approach to detect individual events rather than integrating over many simultaneous ones. Yet, molecular vibrations imply that each electron that passes across the junction will "see" a slightly different species. This effect will be smaller for stiffer molecules and at lower temperatures. Thus, even if transport is not spatially averaged, most observations will still be time-averaged. The assumption of isolation from environment is even more misleading. It is now well recognized that even in the absence of a formal chemical bond (see also section 3.1) the density of states of the system after contacting is different from the superposition of those of the two isolated components,[44-49] a topic that will be considered further in section 6.2.

**2.2.2. Collective effects** are orthogonal to quantized ones, and they provide insight into the emergence of macroscopic charge transport. Given that molecules can interact with their surroundings, monolayers offer a controlled way to manipulate such contributions. Simple examples are dielectric constant and polarization, and others will be discussed in this review. Collective effects are fundamental to biophysics as well as to thin film organic electronics. Considering collective effects is a must from the perspective of technology. Lateral coupling of energy levels also can facilitate transport.[50] Finally, maybe the clearest difference between a monolayer and isolated molecule is the long extension of field lines *around the isolated molecule*, compared to hardly any penetration of field lines *into a molecular array* (2D).[51,52] This leads to distinctly different energy levels at the bottom and top of the 2D film, compared to minor internal polarization for an isolated object[53,54] and 1mJ have pronounced charge migration to/from the substrate, induced by their extended field-lines.[52] The size of the solid contact is also critical: a single Si atom cannot develop image charge in contrast to few-atoms cluster, which already shows such charging.[52] Therefore, monolayers are much more efficient than single molecules in manipulating energy alignment across interfaces, as will be further expanded on in section 6.2.

**2.2.3. Lateral Dimension.** There is a huge difference between a single-molecule junction and a ∼ 10 nm-wide array,[55] but much smaller differences between 10 nm and 10 mm wide junctions.[56,57] Collective effects commonly scale with the nearest-neighbor distance, which is roughly 0.5 nm for linear, non-coiled molecules. Thus a 10 nm wide contact is already ×20 larger than the inter-molecular distance which is apparently sufficient for well-developed collective effects. Further increasing the contact diameter to the 100's of μm range, as is often done[58-62] implies >$10^4$ larger aspect ratios but is not expected to affect the transport physics. Contact diameters in the range of 3 - 300 μm are compatible with photo-lithography or even metallic liquid contacts.[62] A major reservation is that the wider the contact is, the more susceptible it is to defects and irregularities.[62] Indeed, defects are the key drawback of NmJ, as further discussed in section 4.3 below. We also note that the two idealized views of infinitesimally sharp point contacts (1mJ) or perfectly flat substrate (NmJ) are probably misleading. Atomic protrusions, which form the contacts to 1mJ are only a few nm's away from much larger metallic leads. In addition, evidence is accumulating for the critical role that roughness-derived protrusions have in NmJ.[62-66]

**2.2.4.** **Statistical nature / Sensitivity**: From all above-mentioned characteristics it is clear that the soft nature of molecules leads to current measured across a 1mJ to be stochastic, while that across an NmJ is the sum of many electron transfer events. It follows that comparing transport between individual single NmJs is far more reproducible than between 1mJ ones,[19] and far fewer repeats are needed to establish the characteristic transport (10 to 100 scans cf. $10^3$ to $10^4$, respectively).[67] On the down side, large-area averages are often biased[68,69] and dominated by 'hot-spots' of high conductance, while 1mJ transport histograms can reveal a wealth of transport scenarios, reflecting varying details of junction geometry. In principle, it is possible to identify different transport paths within a given 1mJ histogram.[2,70,71] Averaging over large area in NmJ increases the sensitivity of those measurements and thus enables collecting much lower current flux per molecule than in 1mJ. This enhanced sensitivity implies that NmJ are able to measure longer molecules (resistance ~ length, and often exponentially) than 1mJ. Considering that the metallic states can extend into the molecules for few Å, longer molecules can be critical for detecting intrinsic molecular properties rather than hybridization of molecular orbitals with those of the metallic contact(s).[30,45,72] Notice that in practice the nominal current flux per molecule is often (much) higher for 1mJ than for NmJ,[19,26,55,73] an effect that is further discussed in Section 4.3.

**2.2.5.** **Equilibrium state:** Bluntly speaking, 1mJ are inherently in a non-equilibrium state. In the so-called break-junction technique, metallic constrictions are repeatedly formed and broken. In electro-migration, an ill-defined rupture is made by aggressive heating.[4] With the exception of a few beautiful attempts to chemically build up truly isotropic contacts by e.g., C60 dumbbells,[74,75] or covalent binding to graphene reactive edges,[76] 1mJ are not only ill-defined structurally, but they also require relatively weak molecule-electrode binding to enable molecular diffusion into the junction.[77] Both the molecules and the electrodes in 1mJ (except in electro-migration) are most of the time under strong tensile or compressive stress.[78] Especially transport measured using SPM contacts can be sensitive to stretching[70] or pressing[78-83] the molecules on transport. In contrast, judicious design of interactions during and after self-assembly provides an efficient handle to manipulate the molecular configuration within a NmJ. From a technology perspective, the intrinsic stability of NmJ is obviously an advantage.

Alkyl di-thiols for example, serve as a fairly robust standard for NmJ,[28] while their conductance is poorly reproducible in 1mJ,[84] and often show three distinct conduction states attributed to different atomic arrangements of the binding Au atoms.[70] However, even if the probability for meta-stable conformers is lower by orders of magnitudes, the major drawback of NmJ is the inherent uncertainty regarding the possible weight that such low-probability defects have on net transport (see sec. 4.3).

**2.2.6. Experimental handles and complementary tests**: Another fundamental difference between 1mJ and NmJ is the ability to perform complementary characterizations. Considering that a monolayer is in a (near) equilibrium state, and that most spectroscopy tools gain sensitivity by averaging over at least μm-wide detection area, NmJ offers technically feasible complementary characterization by a variety of surface analysis tools, providing spectroscopic information and quality control.[85-90] Single-molecule trapping, though, is mostly blind, and requires a 'transport marker' to identify that a molecule exists in the junction. The most common such marker is a plateau in the conductance-length curve.[30] Further insight can be gained by high-order differentiation (Inelastic Electron Tunnelling Spectroscopy, IETS), noise analysis, and by employing external stimuli other than voltage (transverse electric field), such as electrical gating, thermal gradient or photo-excitation.[24] Such tools are more common for 1mJ than for NmJ, partially because of real technical limitations of NmJ (e.g., gating geometry) and partially due to conceptual fixation. *In situ* photonic activation / recording were specifically demonstrated as technically feasible and highly informative complementary characterization of large-area molecular junctions.[63,64,91-96] IETS can be done in NmJ's[94,97-99] though it requires low temperatures and long-time stability (i.e., impractical with most liquid contacts and common SPM's).

Therefore, array-based (NmJ) and single molecule (1mJ) are two complementary approaches to MolEl, designed to meet different challenges and providing different information to complete our understanding of charge transport in/via the smallest functional material building blocks. The equilibrium, self-assembly of NmJ has some relation to chemical interactions, with applicative prospects; the large area enables complementary chemical characterization to understand the structure-function (transport) relations. While this review clearly focuses on exploring electron transport properties of molecular monolayers, good science will ultimately synthesize all accumulated knowledge, 1mJ and NmJ. In the next section we briefly review the pioneering works in MolEl, all done with NmJs.

## 2.3. Who: Major milestones in measuring charge transport across molecular monolayers

Very often, reviews on molecular electronics do not go back further than the late 1990's, except maybe for the theoretical work of Aviram and Ratner.[11] Choi and Mody wrote an enlightening 'history of science' essay tracing back the origin of 'bottom-up' electronics concepts to the late 1950's.[100] Yet, initial MolEl dates back to the late 1930's, with attempts to measure the electron-ic conductance and capacitance of Langmuir-Blodgett (LB, see section 3.1.3) films of saturated fatty acids, which re-gained interest during the 1960's with the general curiosity about quantum mechanical tunneling (see Ref. 101 for a detailed review of early attempts). Reading through these pioneering papers reveals amazing similarities to contemporary challenges and understandings. Here we give a very brief perspective paying tribute to these ingenious surface science masterpieces. It also serves to introduce the major issues discussed in this review.

The first successful experiment with ~80% yield of reproducible devices was published by Mann & Kuhn in 1971, who measured LB monolayers of Cd-salts of fatty acids on oxidized Al substrate and various top contacts.[101] They extracted a dielectric constant of alkyl-based films between 2.7 to 2.4, which was later confirmed by many follow-up studies on LB films,[102-104] as well as by recent results for thiol-based self-assembled monolayers.[105-107] The tunneling decay coefficient (β, see Eqs. 3, 14 and section 6.3) was 1.45 Å$^{-1}$, somewhat larger than accepted today,[28] though similar to inter-chain values.[108] Mann and Kuhn used a liquid Hg drop as a soft, non-destructive top contact, and compared it to evaporated contacts (Al, Pb & Au), launching the ongoing quest for reliable top-contact deposition[109] (section 4.1 below). The tunneling decay constant varied as the square-root of the metal work-function, in "perfect agreement" with a WKB trapezoidal-barrier model.[101] Although the WKB-description is not popular anymore, the "perfect agreement" tone still prevails, but with other models … .

Honig made a similar study of current transport across 1-15 LB layers of stearic acid (2.5 to 40 nm thick) on different oxides (of Al, Sn and Si) and top-contacted by Hg.[110] He observed 7 orders of magnitude current variation for the different underlying oxides, and a current – voltage dependence that fits Schottky emission at the contact or Poole-Frenkel emission between trap-states in an insulator ($\ln I \propto \sqrt{V}$).[110] The relevant current-voltage description for transport across molecules is still debated, especially at intermediate distances (3 to 30 nm)[29] as further discussed in section 5. Issues of poor reproducibility and identification of good junctions were solved by setting a resistance threshold of 5 kΩ to distinguish a working junction from a shorted one[101] or by a pure capacitive admittance, namely that the ratio of real to imaginary components is smaller than 0.1.[104]

Polymeropoulos extended the study of charge transport across LB films of fatty acids by systematically varying the length of the fatty acids, rather than comparing multilayers, and found a small effect of temperature (down to 170K) on both conductance and capacitance, confirming tunneling as the main transport mechanism,[104] an issue that is still debated (section 5.2 below). The β values were between 1.5 to 1.2/Å, depending on the top-metal contact and overall, the current increased when the monolayer was contacted by metals of lower work-function (Au to Mg). Such dependence on work-function was re-established almost three decades later by conductive-probe AFM work, though with opposite trend.[111] The relation between the work-function of the contacts and the net conductance[43,111-113] is still being explored as detailed in section 6.2.

Polymeropoulos and Sagiv were the first to replace the LB films by what are called nowadays self-assembled monolayers[103] (SAMs, see section 3.1.2), an adsorption strategy that greatly expanded the variety of monomers to be incorporated into monolayers. The deposition method (LB or SAM) did not affect the conductance but changing the head group from a carboxylic acid (fatty acid) to tri-chloro-silane reduced the conductance by about an order of magnitude. They also measured a set of perfluorinated fatty acids which yielded some 4 orders of magnitude smaller conductance than the corresponding fatty acids, and an exceptionally high β value of 2.3/Å.[103] The huge attenuation in conductance might be due to increase in the effective work-function of the Al contact by the close proximity of the highly electro-negative F, which is consistent with hole-dominated transport (Al is mostly a hole conductor [114]). The issue of binding groups re-gained much interest recently as further described in section 6.2. Interestingly, Polymeropoulos and Sagiv also deliberately formed defects in the covalently bonded octadecyl tricholoro silane (OTS) and found that the $\sim$ 5% area perforated SAM has a capacitance similar to that of the homogenous monolayers, but an order of magnitude higher conductance. [103] Defects are further considered in section 4.4.

Roberts et al., were the first to use a semiconductor substrate (InP) rather than a metallic one, in an attempt to avoid the inevitable oxide dielectric that complicates the data analysis. [102] They also studied the frequency response of both conductance and capacitance, and attributed it to either hopping of electronic carriers, or to polarization of the carboxylic acid dipole. [102] The next step was to introduce functionality into the monolayer by using redox-active groups ("electro-chemistry-like" MolEl) or layers of donor-acceptor pairs and dye molecules sandwiched between solid metallic contacts ("organic electronics-like" MolEl). Pioneering work started in the early 1980's, as covered in detail in Ref. [115] (an early review) and more recently in Ref. 1.

We now have reached the "rebirth" of molecular electronics with first reports on *single* molecule junction (1mJ) devices. [67,116-122] This "mesoscopic physics" approach to MolEl is covered in an early review[6] and in several more recent ones. [2,4,5] This review covers NmJs, less covered by recent reviews (for previous ones, see Refs. 19,26,28,123). The widespread availability of surface characterization tools (SPM, XPS) and fabrication tools (lithography, clean rooms and glove boxes), were critical for expanding the research on MolEl (both 1mJ and NmJ). Nonetheless, the fundamental issues that hampered pioneering efforts (reproducibility and reliability, mostly) are often still prominent. Creating a reliable NmJ is not technical details; it is a prime manifestation of bottom-up self-assembly of matter, as considered next.

## 3. Molecular monolayers:

As noted above, the main motivation for studying NmJ is to modify and control the electrical properties of interfaces. In a rather far-off analogy, the importance of an interface can be compared to that of a cell membrane of living matter.[124] It is the cell membrane that preserves a different steady state within the cell than outside it, allowing to treat each cell as a separate entity of a quasi-closed system (quasi-closed because of the transmembrane transport of molecules and ions). The role of interfaces in electronics is conceptually similar. On the one hand an electrical interface ("contact") should allow electrons to cross it (cf. ions and nutrients in a cell) to conserve an electrical equilibrium with the larger system, including the outside; on the other hand an electrically active interface must maintain an electric field across it (cf. concentration gradient / homeostasis in a cell) to induce functionality. A molecular monolayer also shares some chemical similarities with a phospholipid bilayer, or more exactly with half of it, a mono- rather than bi-layer. The major common aspect is the notion of self-assembly or a thermodynamically driven process that can lead to perfect ordering and effective functionality.[125] Mild interactions and some structural flexibility are other characteristics common to both bio-membranes and artificial monolayers, though these aspects can vary considerably between different types of monolayers.

A rich literature exists on monolayer formation and properties.[16,126-131] Monolayers were studied for fundamental understanding of aggregation and crystallization[132] and toward their use for modifying surface properties like corrosion and diffusion inhibitors, friction, wetting and directed growth of overlayer films, adhesion promoters / inhibitors and bio-recognition.[129] Here we briefly cover only those aspects in monolayer formation and properties that are relevant to electrical aspects. The next sub-section compares different strategies to form monolayers (3.1) and design guidelines in the context of electronics (3.2); we then describe characterization of monolayers both structurally (3.3) and electrically (3.4).

## 3.1. Three generic types of monolayers

Molecular monolayers and electrical contacts to them, are often classified according to the binding strength between molecules and the electrodes,[26,28] so-called 'chemisorption' and 'physisorption'. There is no sharp distinction between these two,[133] an issue that is closely related to the problem of sharply defining the concept of a chemical bond.[134,135] (see also sections 4.2.3 and 6.2, below). It is clear though that the chemical forces, which drive the adsorption process, are crucial for dictating the monolayer structure, uniformity and long-term stability. Three typical modes are shown schematically in Fig. 1, and briefly described below. Here we focus on classifying the different methods, their strengths and limitations and the interested reader is referred to dedicated reviews for a detailed overview of the richness and diversity of methods to form molecular monolayers.

### 3.1.1. Non-specific adsorption;

Non-specific adsorption is most common for vacuum deposition of molecules; it can be used for a huge variety of molecules from small ones (gaseous at STP) up to large poly-aromatic ones. Such depositions are often not self-terminating after a single layer is completed and the thick-ness is limited by dosing. The molecule – substrate attraction should be stronger than the molecule – molecule one ('good wetting') to get uniform coverage rather than multi-layer islands. The 'non-specificity' refers to the fact that the adsorbate does not include a specific binding site; however, often the adsorbates have a preferred orientation with respect to the substrate's surface. The dominant interactions are those between the adsorbate and the substrate, such as van der Waals ones, interaction of π orbitals with metal atoms (especially relevant for organic conductors) and even charge exchange (see section 2.2.1 above and 6.2 below).[47] Inter-molecular interactions can induce lateral ordering (often studied by STM), but are generally much weaker than the interaction with the substrate. Temperature and dosing are the main experimental handles controlling the monolayer formation. Non-specific adsorption prevails in thin-film organic electronics, STM studies of molecules and in some single-molecule MolEl.[136]

### 3.1.2. Self-assembled monolayers:

The key to the mechanism of self-assembly is the addition of a specific binding group to the molecular skeleton (shown in Fig. 1 by yellow and purple moieties, respectively), such as silanes (binding to glass)[137] or thiols (binding to gold).[138,139] The binding group, also known as 'head' or 'sticky end' has a much stronger affinity to the substrate than the rest of the molecule, also known as 'tail'. The tails play a critical role in dictating the net order in the monolayer, by the strength of their mutual interactions, such as van der Waals (alkanes)[140,141] or π stacking (aromatics),[142] as well as other interactions that are specific to certain chemical groups, such as H-bonding[143,144] or Coulombic repulsion.[145,146] These tail interactions are rather weak and therefore extremely sensitive to details of the binding reaction, such as temperature, humidity[147] or concentration.[148] Mutual repulsion between head-groups occurs in cases of charged (e.g., phospho-nate)[86] or bulky head groups. Commensuration between molecular footprint and surface bind-ing sites is also critical in dictating net monolayer structure.[140,142]

In MolEl, reactive groups are often added at both ends of the skeleton (e.g., di- instead of mono-thiols[149]) to enable binding to both electrodes.[150-153] Such dual binding is common practice in break junction MolEl.[30] However, for monolayers special care should be taken to avoid 'looping', i.e., avoid that both reactive ends bind to the substrate, or mutually bind (e.g., two R-SH form R-S-S-R) to yield a bi- or multi-layer.[148,151,154,155] Note, though, that multilayers of controlled thick-ness can well be a desired end product, which can be achieved if the molecules are terminated by groups with attractive interactions *with each other*.[156-158]

Self-assembled monolayers are most often formed by immersion in a solution of the monomers, but can also form from the gas phase or from neat melted monomers. Strictly speaking, 'self-assembly' implies that no additional stimulus is needed to induce the binding; however, this term is commonly used to describe monolayer formation by binding heads, including cases where the binding is induced by heat, light or electrochemistry. This latter method often leads to stronger, covalent binding to the substrate.

SAM formation is the main procedure used for making monolayers for NmJ MolEl. The interest-ed reader is directed to recent reviews on monolayer formation in general[131,159] and onto specif-ic substrates, such as metals,[129,160] oxides[161,162] or oxide-free Si.[16,88,163,164]

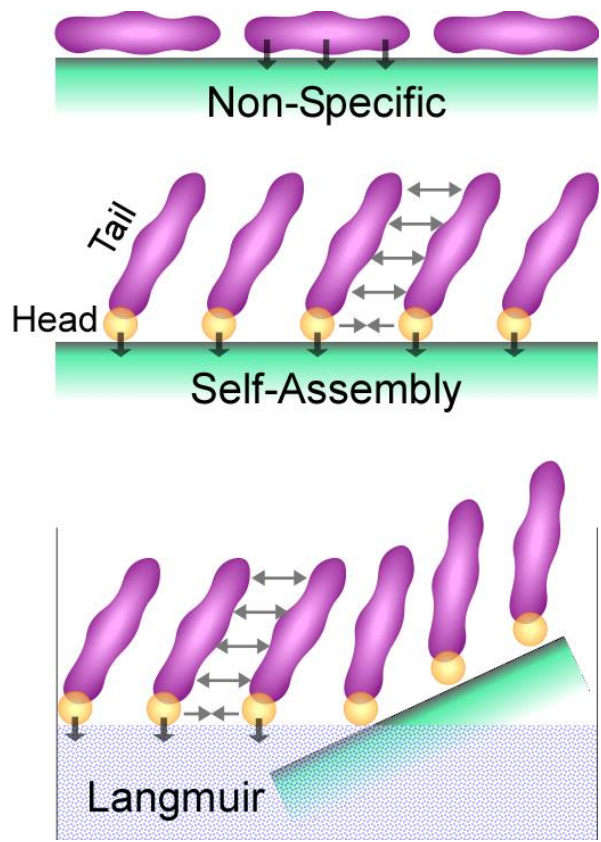

**Figure 1:** Three archetypical monolayer adsorption procedures. The purple part represents the tail or the main portion of the molecular, while the yellow circle stands for the head or 'sticky end'. The vanishing green is the top surface of an ~ infinite substrate, and the bluish shade in the bottom scheme represents the aqueous side of an air / water interface. Arrows indicate major interactions.

### 3.1.3. Langmuir trough-based monolayers:

This class of monolayers is the closest to biological membranes, and is the origin of the head / tail concept later used for self-assembled monolayers. The major difference is that formation of 'self-assembled monolayers' is driven by adsorption to the substrate, while Langmuir monolayers form at the liquid / gas (or other liquid) interface. This implies that the head and tail distinction of Langmuir films is mostly in terms of hydro-phobicity / -philicity (i.e., amphiphilic molecules) rather than chemical reactivity. Langmuir films are often known as Langmuir-Blodgett (LB) ones, a term describing the method of transferring them onto a solid. A different transferring method is Langmuir-Schaefer, though often this distinction is overlooked and the acronym LB is used for transferred films in general, as is also done here.

The pioneering experiments in MolEl were done on LB films.[101,103] For MolEl this technique is inferior to self-assembly in two aspects: weak, unstable binding to the substrate and ill-controlled ionized residues (such as water molecules) and partial substrate oxidation. Still, conceptually it is an interesting alternative. Langmuir monolayers are the extreme opposite case to covalent binding, with ultimate lateral diffusion of monomers along the interface. This allows the system to exhibit different phases in terms of molecular arrangement. It will be fascinating if a similar control over surface arrangement can be obtained also with more stable monolayers, toward study of the collective effects mentioned above (section 2.2.2).

The head groups in Langmuir monolayers are most often carboxylic or phosphonic acids. The binding groups used for self-assembly (e.g., thiols, amines, C=C) are not sufficiently polar to make the molecules amphiphilic. Yet, polar head-groups can induce amphiphilic ordering also at a solvent / solid interface and are not limited to an aqueous phase as in a traditional Langmuir trough. For example, if a very non-polar solvent is used (e.g., chloroform) polar head groups are repelled from the solvent and form ordered molecular arrays at the surface of the solvent. This principle was used to form monolayers by the so-called T-BAG technique[165] and also by just spin-coating.[166,167] These methods do not involve water, which is a marked advantage over some other methods (minimizes substrate oxidation and ionic residues) and if the substrate is hydroxyl-terminated (as in e.g., oxides) the acidic head groups can further react by condensation with the substrate to establish stable binding. Overall, polar head groups offer a unique design motif, which seems not yet to have been fully exploited.

## 3.2. Structural Considerations in monolayer preparation

### 3.2.1. Choice of Substrate and binding chemistry

The substrate is also critical in dictating the net monolayer performance. Hydroxyl-terminated metal oxides are very attractive for their facile binding chemistry, but much less so from the electronic perspective. Most oxides are insulators and therefore their thickness has to be limited to only a few nm. Such thin oxide films are rarely stoichiometric and have a high density of charged traps. In this regard the Si-H termination is special because, on the one hand it is sufficiently stable to protect the substrate from oxidation for some 10-30 minutes under normal ambient conditions, and on the other hand the Si-H bond is weak enough to be replaced by the desired stronger chemical binding.[16,88,163,164] Si-Cl can behave similarly[87,168] and it was shown that Cl-termination of III-V semiconductors can also induce efficient binding of alkyl monolayers that keep the substrate oxide-free.[169] An intermediate approach is to use atomic layer deposition, ALD, to deposit a minimal thickness of a well-controlled oxide, such as $Al_2O_3$, on a Si substrate and use the oxide's surface hydroxyls to further bind acidic head groups.[170] While oxides[162] and semiconductors[16] yield more stable monolayers than noble metal substrates, from the electrical transport point of view they have a major drawback of complicated, up to ill-defined electrical conductance, which considerably hampers our ability to reliably interpret the measured transport properties.

Covalent bonds are also formed between *in situ*, electrochemically made radicals and several types of electrodes, C,[158] Si[171] or Au[172]. Diffusive binding to carbon electrodes is also possible via π stacking with poly-aromatic moieties.[75,173] Surface diffusion is prominent in the case of thiol on gold, were the RS-Au bond has a similar strength to the Au-Au bond of a gold surface atom to the lattice.[77] This bonded Au atom is known as an 'adatom' and the molecule + adatom adduct diffuses on the surface to reach the thermodynamically preferable position for the *tail*. This mechanism leads to highly ordered and even crystalline monolayers of alkyl-thiols on Au,[174] but is also responsible for formation of 'etch pits'[175] considerable re-shuffling of Au surface atoms[77] and grain boundaries (see below).

Silane and phosphonate binding to hydroxyl-terminated surfaces present an intermediate case, where the binding occurs in two steps. In the first step, the heads are weakly adsorbed to the substrate (often they are actually repelled out of the solvent) and at a later stage, a condensation reaction leads to covalent binding.[165] In the case of silane much of the covalent binding is between head-groups and only a limited amount of binding occurs to the substrate.[176-178]

### 3.2.2. Order and dense packing

The reactivity of the binding group plays a critical role in dictating the order and stability of the monolayer. Weak binding groups facilitate surface diffusion of the monomers and, therefore, favour ordering in the monolayer. The lattice-spacing of the substrate is important as it often imposes the binding distance between monomers.[140,179] The mutual (van der Waals) attraction of alkyl chains ("tails" of the alkyl molecules) increases with their length, and monolayers of alkyl chains are considered to be ordered only above ~12 carbons (~1 nm).

A disadvantage of such order in a monolayer is the formation of grain boundaries between highly-ordered domains. Thus molecular junctions with a contact area larger than the monolayer domain size will probably be dominated by shorts across defects at the grain-boundaries.[62] Monolayers on rough surfaces,[180] or composed of molecules with large head-groups[86] are more prone to defects. In such cases, short-tailed, less-ordered monolayers allow for better healing of defects in the monolayer than long-tailed, ordered ones.[86,180] Close packing in a monolayer is expected to restrict the motional freedom of the molecules.[181,182] For this reason, photo-isomerization switching is often less efficient in a monolayer (≥30 min for completion)[76,183,184] than in solution or in 1mJ (∼sec).[185]

The binding strength is also expected to affect the ability of the monolayer to withstand subsequent treatments, mainly top contact deposition, as stronger binding to the substrate is expected to help sustain harsh physical or chemical treatments.[158] Strong Inter-molecular interactions also contribute to the endurance of the monolayer.[143,144,186] Elimination of defects is critical not only for electrical transport across them, but also to prevent diffusion through the monolayer of small molecules (e.g., $O_2$ or evaporated metal atoms) that alter the surface chemistry.

Diffusion in the substrate is a parameter that cannot be underestimated as a source of artefacts in molecular electronics. Some of the pioneering 'molecular switches' were later recognized as diffusion of oxygen vacancies within $TiO_2$[187] or $SiO_2$[188] matrices, or electro-migrated gold stalactites.[65,189] The inorganic effects were reproducible and technologically relevant for memristor applications, though they were not 'molecular'. These examples convey the message, that the utmost (z-direction) miniaturization required by MolEl leads to a range of novel phenomena not necessarily originating at the molecules.

### 3.2.3. Introducing advanced functionalities

The requirements for order, stability and functionality often are not compatible. In addition one has to worry about self-reactions between the sticky and functional ends for large enough molecules.[16] An accepted route to optimize between these contradictory needs is to do *in situ* a follow-up reaction on the bonded monolayer to stabilize it, varying from metal-ligation[16,190] condensation[191] to 'click' chemistry.[16,192,193] The last option is especially attractive as it concerns a reaction under mild conditions with minimal degradation of the monolayer, lack of reaction residues that can be trapped in the monolayer (though catalysts can still be trapped and traces of Cu on Si substrates prevent use of that catalyst with Si substrates for MolEl). Another strategy is mixed monolayers, where electrically active monomers are embedded in a generally inert monolayer (e.g., alkyls).[20,117,193-195]

## 3.3. Structural characterization of monolayers

A clear advantage of monolayer-based over single molecule-based MolEl is that we can obtain plenty of structural information on the monolayers and ideally even have practical handles to manipulate the structure (section 3.1), and by that learn about the relation between structure and functionality. Characterization of monolayers is often much more technically demanding than their preparation. Their nm-scale thickness generally pushes any characterization tool to its sensitivity limit. The need for sensitivity often translates into a need for large area samples of 100s μm to mm diameter. In contrast to most surface science, which is induced and studied within the same vacuum chamber, wet-surface modifications are done outside the characterization chamber. The environmental stability of these samples, over time and under different environments (e.g., humidity, dust, nature and pressure of ambient atmosphere) cannot be over-estimated. Artifacts and un-intended ("adventitious") adsorbates can be easily mistaken as molecular effects. In a 2010 review we elaborated on surface analysis tools relevant for monolayer characterization.[85] To make the present review self-contained, we briefly note the main ones.

### 3.3.1. Thickness and density

Monolayer **thickness** and / or **density** can be deduced from optical measurements by **ellipsometry**,[196] from photoemission data by X-ray photo-electron spectroscopy (**XPS**)[197] or from X-ray reflectivity (**XRR**),[86,198] in increasing order of complexity and accuracy. Ellipsometry is appealing as it can be done in ambient (and can thus include contaminations that might also be there on the actual sample if it is measured in the same ambient atmosphere) and is the least destructive; thus, it is often used for quick quality control and not necessarily as a reliable quantitative method.

Determination of thickness is convoluted by the uncertainty about the density (which affects the refractive index in ellipsometry, the inelastic mean free path in XPS) and roughness (i.e., non-uniform layer profile). However, the practical assumption is that the refractive index or inelastic mean free path are fairly insensitive to the details of the surface film. While *spectroscopic*-ellipsometry is widely used, it is *angle*-resolved ellipsometry that is more informative about the refractive index and the surface profile (e.g., distinguishing between an organic layer and a thin surface oxide onto which the former is deposited).

XRR is the only method that can extract all these characteristics (thickness, density and roughness, also for a multilayer sample) independently, but it relies on complicated fitting which requires very high quality spectra, meaning long collection time (up to 12 hours) and homogenous large area substrates ($\sim 4$ cm long). Binding **density** is often inferred from combining ellipsometry thickness with XPS attenuation,[199] but this is a convoluted determination which reduces its reliability. Atomic force microscopy (**AFM**) is also used for measuring thickness by first scratching through the monolayer and then re-scanning.

**Homogeneity** and **roughness** are often studied by AFM, though qualitative, rapid information can be gained also from **contact angle measurements**, preferably with various liquids beyond water, and by comparing both advancing and receding droplets.[200-202]

### 3.3.2. Chemical composition

Evidence for the **chemical nature** of the surface modification is critical for establishing the correct surface modification. Here the main tools are **XPS**[203] and (Fourier-Transform) **infra-red spectroscopy (FT-IR)**.[88,204] XPS gives (near) surface composition and charge state while IR, over the 600 to 4000 cm$^{-1}$ range, can provide information about the presence and orientation of specific organic groups (e.g., hydroxyls, methylene orientation, gauche defects, amide bond, binding to inorganic substrate atoms). **Raman** spectroscopy is very popular for 2D materials, like graphene, but less used for organic molecular monolayers due to their low absorbance. Surface roughening to induce plasmon-enhanced Raman easily leads to electrical shorts (though see Refs. 92,205 for exceptions).

Direct IR absorbance can be sufficiently strong if large apertures are used, but often special accessories are used to enhance the signal to noise ratio, such as attenuated total reflectance (**ATR**) or IR reflection absorbance (**IRRAS**)[204] from metallic substrates. If the substrate is transparent over the relevant IR range (which is the case for most semiconductors) it can serve as a waveguide after polishing its facets in a mode known as multiple internal reflection (**MIR**).[88] Pressing a reflecting sample against an external ATR crystal, especially a Ge one, gives very high sensitivity, with minimal preparation work (though the matching of the refractive indices of sample and ATR crystal should be considered). The signal to noise ratio of IRRAS can be further enhanced by using polarization modulation (PM-IRRAS), at the price of severe skewing of the spectrum, rendering its intensity less quantitative.

Advanced photo-electron spectroscopy can be done in a synchrotron facility, which enables scanning the energy of the exciting photons, as in Near Edge X-ray Absorption Fine Structure (**NEXAFS**) which is informative on molecular orientation.[204] Secondary ion mass spectrometry (**SIMS**) is informative on chemically bonded fractions and can detect the formation of metal-organic species,[206,207] and provides high contrast chemical mapping of a surface.[208] SIMS is one of the most sensitive surface characterizations and therefore can be used to elucidate minute degradation routes, e.g., penetration of top-contact.[206,207]

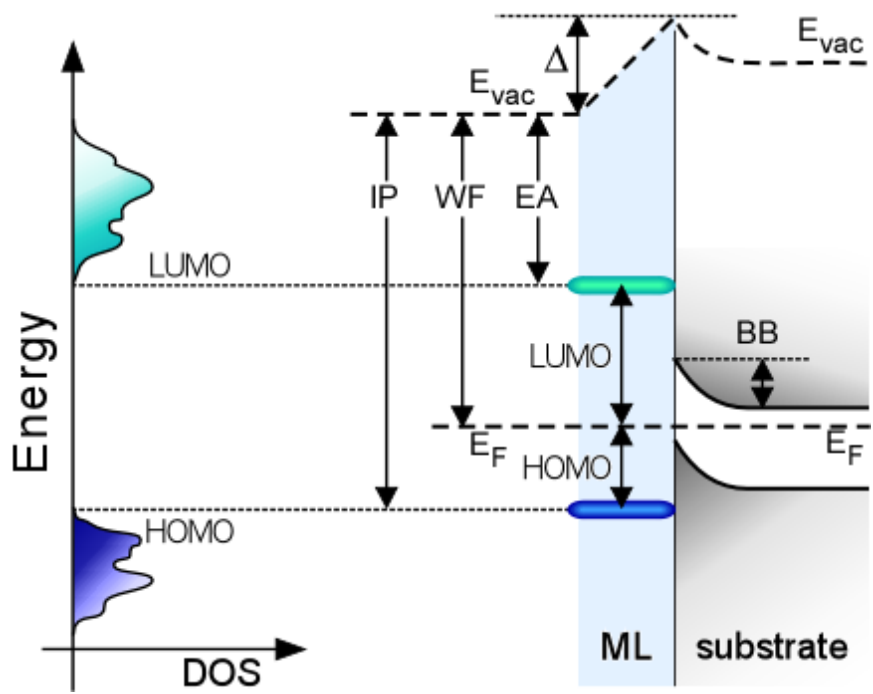

**Figure 2**: RIGHT: One electron energy diagram illustrating the major electrical characteristics of a molecular monolayer (light blue region; right) adsorbed on a solid substrate (gray region; right). 'HOMO' and 'LUMO' indicate both the energy level and the difference in energy between these levels and the Fermi level ($E_F$) of the substrate; the ionization potential (IP) and electron affinity (EA) are the energies measured from the vacuum level ($E_{vac}$). The substrate is a semiconductor, characterized by a forbidden energy gap (white band) and surface band bending (BB); a metallic substrate is similar except that the Fermi level resides within a continuum of allowed states and no BB exists. LEFT: a more realistic diagram of the density of states (DOS, horizontal coordinate) as a function of energy (vertical) near the Fermi level. The frontier levels of adsorbed molecules are significantly broadened.

## 3.4. Electrical characterization of monolayers

While charge transport across interfaces is an ultimate goal for most of MolEl, there is a variety of non-transport tools to characterize electrical properties of modified surfaces. There are four key electrical properties of surfaces, as illustrated in Fig. 2:

1) the energy difference between the Fermi level and the local vacuum level, its **work-function (WF)**;

2) the macroscopic potential drop near the surface, characteristic of non-metallic solids, known as '**band-bending' (BB);**

3) the energy levels, closest in energy to the Fermi level, i.e., for the filled states the valence band for extended solids or highest occupied molecular orbital, **HOMO,** for molecules, and

4) the empty states, conduction band for extended solids or lowest unoccupied molecular orbital, **LUMO** for molecules.

HOMO and LUMO are generally reported with respect to the Fermi level of the system while their distance from the vacuum level (the energy to remove or add an electron) is known as the ionization potential (**IP**) and electron affinity (**EA**), respectively. The energy levels of molecular films are much broader than those of isolated molecules (e.g., in gas phase). It results a continuum of states as illustrated on the left side of Fig. 2 in contrast to isolated level as is often considered in simplified models (right-side of Fig. 2 and Fig. 6 below). These properties are critical to any electronically active interfaces, and comprise an essential part of the study on novel electronic materials.[8,209]

In the context of molecular electronics, the alignment of the molecular energy levels with respect to those of the electrodes (one of which, for NmJ, is the substrate) is critical for their net charge transport. Furthermore, molecular adsorption can alter the molecular energy levels by hybridization and creation of new energy levels or, less drastically, by partial charge rearrangement between the adsorbate and the substrate, leading to changes in the net work-function and band bending. Thus no genuine understanding of the molecular role on charge transport can be gained without this complementary characterization. Yet, an important caution is that these techniques are generally limited to surfaces and cannot detect the energetics at interfaces, buried under another solid phase.

The two key techniques to study surface energetics are Kelvin-probe measurements and photo-electron spectroscopies. The **Kelvin probe** method is a capacitive, non-contact one that measures the contact potential difference (**CPD**) between a metallic probe and a sample, which is basically the difference between their work-functions.[210] The probe is brought into close proximity ($\sim$100's μm) to the sample to form a capacitor between them, with its net charge proportional to the CPD, which is the readout of the technique. Both macroscopic ($0.1 - 2$ mm wide) and nm-sized (for AFM modification) probes can be used. The measurement can be done in any environment as long as there is no contamination of probe or sample surface. CPD is often reported as a relative measurement; translating it into an absolute value of the work-function of a given sample, requires pre-calibrating the probe against a stable reference standard, such as freshly-cleaved HOPG.

Photo-electron spectroscopy (PES) is a more versatile technique to determine surface energetics (Ref. 8 reviews PES in the context of molecular electronics, and Ref. 209 provides a didactic introduction). The focus on low binding energies, and desire for surface-specific sensitivity, requires UV photons rather than X-rays (UPS cf. XPS, respectively). Both UPS and XPS are by definition sensitive to filled states only, relative to the position of the Fermi level (=0 binding energy). The onset of the first peak is identified with the HOMO position or valence band maximum (see left side of Fig. 2). The difference between the energy of the incident photons and the largest measurable binding energy of photo-excited electrons (known as 'cut-off' energy) is the work function of the sample.

The empty, LUMO levels are highly relevant for the net electronic performance. Yet, technically their measurement is much more demanding and practiced less often. It can be done by two photon photo-electron spectroscopy (**2PPE**),[8,211] inverse photo-emission spectroscopy (**IPES**),[209] and somewhat by NEXAFS.[212] Similar to the onset of the filled DOS (HOMO position) it is critical to know the onset of the empty DOS (LUMO), as the difference between them is the energy gap, relevant for transport. In principle, measuring the optical gap (by e.g., **UV-vis** absorption) can be used to estimate the LUMO position without the need for advanced PES, but this is only valid if the exciton binding energy is negligible. However, the low optical absorbance of monolayers, combined with that they are often on an opaque substrate, makes this difficult with standard spectrometers (see Refs. 184,213 for two of the exceptions).

The evolution of surface energetics with film thickness is often studied using in-situ film growth.[214] This is impractical for solution-adsorbed monolayers. Instead, some insight into the potential profile across the monolayer can be gained by a chemically resolved electrical meas-urement, **CREM**.[215] It is based on the ability of XPS to detect each of the elements separately; it monitors the shift in their binding energy induced by a build-up of a static potential across the monolayer. Knowing the chemical structure of the monomers, each element is associated with a different position across the interface, to construct a full potential profile. The CREM method elucidates microscopic resistance variation along the molecule.

### 3.5. SAMs compared to other nm-thin films

In terms of electrical properties and interface modification, there are no fundamental differences between organic monolayers and other, inorganic-based tools to modify interfaces. While there is much research on inorganic insulators and 2D materials, molecular monolayers might offer some unique advantages, especially for low-cost, large-area processing. For the sake of completeness we provide a short bench-marking to major alternatives, though no attempt is made to cover these fully or evenly.

The key features of molecular monolayers are

a) preparation from liquid solution at ambient temperatures

b) closed-shell stoichiometry, i.e., they can be chemically unreactive.

This immediately defines their strengths and weaknesses. Molecules provide high confidence in the composition vertical to the interface, yet, an inevitable high uncertainty in the presence of microscopic defects or voids, at different lateral positions.

*Molecular beam epitaxy (MBE)* probably offers the most reliable structural control (and highest production costs), and is often used for interfacial energy alignment.[216,217] The major limitation of MBE is the need for lattice matching between the substrate and the deposited layer. The notion of epitaxy is important, since without it, irregular islands will nucleate and the result will not function as an interfacial layer, separating two phases. Naturally, the epitaxy /lattice matching condition does not apply to amorphous layers, such as certain oxides.

*ALD* provides high-quality films at much lower cost than MBE. ALD-grown oxides are superior to those grown by e.g., sputtering or thermally, in terms of (smaller) roughness and control of trap density; combined with self-assembly of organic monolayers they offer a powerful composite of inorganic and organic thin layers.[218]

*Thin oxide-films* are the workhorse of the CMOS industry. The major limitation is the tendency to form sub-oxides for very thin film thickness (< 3 nm), due to diffusion into the substrate and equilibration with ambient $O_2$. These structural defects are the major source for leakage currents across oxides, and charged traps. Organic monolayers can in principle overcome these limitations,[86,170,219] but achieving monolayers of comparable quality is a very tough challenge.

While oxide films can be good diffusion barriers (see section 4.2.1, below) a *disadvantage* is their open shell at the surface, which makes them *chemically reactive*. A dense alkyl monolayer blocks surface reactions quite efficiently,[220] compared to oxides, which often react in various aqueous solutions. A 1.5 nm thick alkyl monolayer can protect a Si wafer from oxidation even after one hour in boiling water.[221]

In summary, we can view molecular monolayers as a special case of the various possibilities to modify interfaces. Although this review focuses on such monolayers as a unifying theme, the underlying principles are probably common to a wider range of thin-films.

## 4. Making a junction: Inserting a SAM between two solid electrodes

The sensitivity of the majority of organic molecules to heat and high energies excludes most physical vapour deposition methods, common in standard micro-electronics fabrication. Similarly, standard patterning by lithography is based on organic photo-resists, which are deposited and lifted using organic solvents that can damage/change organic films.[222] Furthermore, since the measured objects, molecules, are the size of common surface roughness or contaminations, it is extremely difficult to control the exact composition and uniformity of the interface, and, maybe more seriously, to get experimental information on such. As noted above, single-molecule junctions (1mJ) are structurally not better defined than large-area ones (NmJ), though the structural diversity is better manifested by recording each event individually. The massive averaging inherent to large-area molecular junctions (NmJ) requires serious efforts to make reliable, permanent contacts to such ultra-thin layers. This challenge is well acknowledged[28,109,158,223-225] and has spawned plenty of creative solutions. Still, none of them has been accepted as a universal wiring method and maybe more severe is the lack of a reliable procedure to assess the presence of defects or hot-spots and their contribution to net transport. These two aspects are covered below in sub-sections (4.2) and (4.3), respectively. We first start with a short classification of the available strategies.

## 4.1. Variety of contacting schemes

As summarized above, large-area molecular junctions (NmJ) were produced already during the 1960's, and in 1971, Mann and Kuhn were the first (of many to come) to claim that their molecular junctions are reliable.[101] Nonetheless, as noted by Metzger, MolEl is a "perennial adolescent",[1] for almost half a century. While recent progress, using different modes of scanning probe microscopy, (for 1mJ or for 'few molecule junctions') is remarkable, these studies are rarely concerned with reliability or permanent constructions (such as obtained using, e.g., evaporated contacts or by electro-migration, in contrast to temporary contacts like Hg, In-Ga or break-junctions, see 4.1.2 below). The biggest issue is the challenge of wiring nano-metric objects to the outside world. The intrinsic difficulties are fragility and reactivity of the organic molecules, diffusion through the layer, and the surface roughness with dimensions similar to, or even larger than the tested items. While apparently technical, this is actually a fundamental challenge touching the heart of 'bottom-up' nanotechnology.[100,125] Can we devise reliable strategies to direct the building blocks of matter? Notice also, that reliable NmJ fabrication is only the first step toward 'bottom-up' construction of intricate architectures made of different units,[226] which is a far larger challenge.[125] Below we outline generic alternatives / choices among a large diversity of wiring approaches, to construct molecule-based devices.

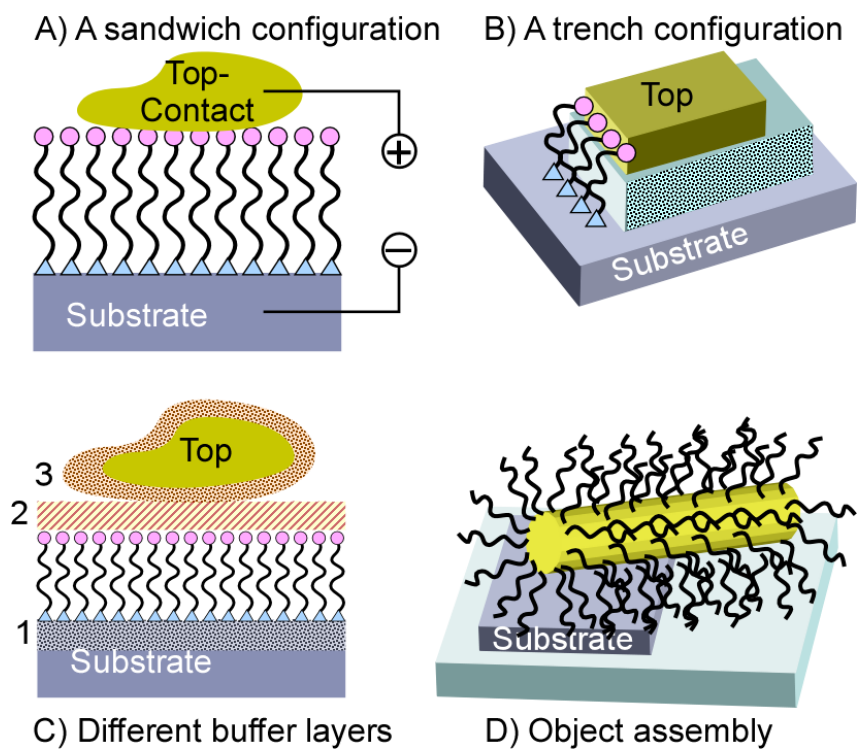

**Figure 3:** Schematic prototypes of array molecular junctions (NmJ, see titles in figure). Wiggly lines stand for molecules, anchored via a binding group (blue triangle) to the substrate (gray box) and with a terminal group (pink circle), pointing toward the top contact (yellowish). Dashed / dotted regions in (C) stand for oxide or interfacial layers, marked 1-3 (see text). In (D) the molecules are adsorbed to a nano wire.

### 4.1.1. Sandwich / Trench

A 'sandwich' configuration has the merit of using the molecules as a ruler, dictating the spacing between the electrodes.[227] They are made in two steps: first, a molecular monolayer (wiggly lines in Fig. 3) is formed on a 'bottom electrode' (grey box in Fig. 3A), such as a silicon wafer or thin film of evaporated metal on glass, and later, a 'top-contact' (yellowish box in Fig. 3) is placed on top of the SAM. Pioneering works on MolEl used such a sandwich approach,[101,103] which remains the main testbed of NmJ.

The major disadvantage of the sandwich approach is the extent to which the molecules are affected / damaged, upon top contact deposition. 'Trench' is an alternative approach where both electrodes are deposited and patterned at the first stage (Fig. 3B), and later the molecules are adsorbed to both electrodes simultaneously.[228-234] The gap or the trench between the two electrodes is often parallel to the substrate and is prepared by advanced lithography, or electromigration.[228,229] The trench can also be vertical (see Fig. 3B) where a thin insulating layer (e.g., $SiO_2$, SiN) defines the opening of the gap.[230-233] An important advantage of either trench configuration is its 'open geometry', enabling gating or concurrent spectroscopy.[92,93] However, the insulator also adds a major problem of leakage current. Thus, the tested molecule(s) should conduct considerably better than the supporting insulator. In addition, fabricating gaps in the order of 1-2 nm is at the very edge of current technology. Thus, trench-like junctions normally require longer molecules, though electro-migration[228] can make narrower gaps, and are the preferred technique for so called single electron transistors (SET).[229]

## 4.1.2. Temporary vs. permanent contacts

Returning to the sandwich configuration, its major limitation is damage induced by depositing the top contact. A possible remedy is using temporal contacts, which are thought to be a) less destructive to the molecules, and b) easier to use for multiple repeat measurements than permanent ones. By definition, temporal contacts are aimed at lab use only. Scanning probe techniques are obviously temporal. While they are often used for 1mJ, they can also be used to contact stable monolayers, resulting in a junction somewhat intermediate between 1mJ and NmJ.[82,235-237]

Metallic liquids are widely used as temporal contacts. The high surface tension and semi-noble character of Hg made it an ideal candidate for contacting monolayers; however Hg is reactive toward common substrate materials (Au, Ag, II-VI semiconductors). If amalgamation would happen only upon direct solid/liquid contact this would not be a serious issue, because of the high surface tension of liquid Hg. However, the reaction can occur with Hg vapour and, unfortunately, the vapour pressure of Hg at room temperature is very high (for a metal). Thus, amalgamation strongly limits its usefulness for contacting monolayers made on these popular substrates (see also section 4.2.1 below). Hg works well for monolayers adsorbed on oxides,[110,238,239] or oxide-free silicon.[239-242] The Hg reactivity can be partially solved by protecting the Hg by adsorbing a thiol-linked SAM directly on it, and measure the charge transport toward a bare substrate,[238,243] or toward a monolayer-covered substrate, to form a bilayer junction.[105,244,245] In the last decade however, SAMs on noble metals are often contacted by liquid InGa alloy (also known as eutectic GaIn or EGaIn). In-Ga alloys have many orders of magnitude lower vapour pressure than Hg, are not or less reactive towards most of the substrates, but are covered by an ill-defined Ga-oxide film,[246] which complicates modelling results quantitatively.

Generally, the fluidity of liquid metal contacts is a drawback for miniaturization, and for variable temperature measurements, because of change in volume/contact area with temperature. Some of these limitations can be addressed by micro-fluidic channels, which deliver and pattern the liquid contacts (mostly InGa),[247] and liquid metals within microfluidics are even studied as electro-mechanical systems.[248] Microelectromechanical systems (MEMS) are also used as gentle temporal contacts, though rarely. A magnetic field can be used to bent a narrow wire and bring it into contact with a monolayer-modified opposing wire.[249]

Permanent sandwich contacts are most often made by evaporation of a metallic film, mainly Al or Au. Although seminal contributions to MolEl employed evaporated top-contacts,[101,103,250-253] the approach was severely criticized because of the damage caused to the SAM, as indicated by IR and TOF-SIMS studies, by the highly energetic impinging atoms.[223,254,255] Sample heating (thermal source) and/or electron irradiation[256] (e-gun source) can also cause damage to the monolayer during physical deposition, such as molecular fracture, and de-protonation to cause cross-linking or unsaturation (see Fig. 4A).

Attempts to reduce the evaporation damage included use of ultra-small contact areas (*e.g.*, "nano-pore",[250,253] shadow evaporation,[257] or nano "crossbar"[258]) to facilitate strain release at the periphery of the junction, but the yield of working devices was rather low,[68] with the inevitable difficulty of selecting the 'genuine' devices. Another approach is reducing the impinging energy of evaporated atoms, by various means. The simplest ways are using an e-beam rather than heat to vaporize the source metal, cooling of the sample, large source-sample distance, very slow deposition rate or shuttering the atomic flux (these 'tricks' are often not transparently reported). In underline{indirect evaporation} the sample is back-faced to the source, and the evaporated atoms reach the substrate by scattering from back-filled Ar.[259-262] Other variations are underline{shadow} evaporation near the molecules, followed by <u>diffusion</u> onto the molecules[263] and covering the SAM by a condensed Xe layer during evaporation to buffer between the SAM and the impinging atoms. The Xe evaporates when the sample is heated to room temperature.[264] Similarly, a non-intended condensed water layer is at times present when evaporating on a cooled substrate, in standard evaporators, especially in the absence of a cold finger inside the chamber that is held at a temperature below that of the cooled substrate. Evaporating in ultra-high vacuum can avoid water condensation and also generally reduce the incorporation of contaminants and interface oxidation toward overall higher reproducibility and yield of working devices.

### 4.1.3. Object assembly / Atomic deposition

Depositing the top contact is not limited to evaporation. ALD offers a relatively gentle atomic level deposition, where organic precursors are carried by a gas flow. ALD requires wetting of the substrate by the precursors, and therefore mostly requires specific terminal groups, like COOH.[265] Other soft deposition methods are chemical bath deposition (CBD) or electroless deposition (ELD) by chemical reduction of metal ions.[181,266-268] By using a careful choice of bath conditions Walker and co-workers were able to deposit Ni and Cu on thiol monolayers without seeding by foreign elements (e.g., Sn or Pd) and control selectivity to specific terminal groups.[268,269] More importantly they found ELD conditions which prevent the penetration of metal through the monolayer. ELD and CBD are also beneficial because of the rather low temperature they are using (up to 40-60 °C), which decreases the probability of molecular degradation.[268,269] However, the driving force for ELD and CBD is extreme pH conditions that can also harm the monolayer[266] and as in any solution-based deposition, co-deposition of residuals is always a worry.

Nonetheless, any such film-growing involves by definition, high-energy species regardless of whether it is high kinetic energy and radiation (evaporation), chemical reactivity (ALD) or extreme pH conditions (bath deposition). In contrast, if the two solid contacts are pre-defined (as in the above-mentioned trench configuration) the system is much less reactive and the only driving force for its assembly is the chemical affinity of the molecular terminals to both electrodes. The trench' disadvantage of large, pre-defined gap can be overcome by transferring small metallic objects, such as nano-wires or thin, floating films onto the monolayer. The fact that these objects are solid and not isolated atoms or reactive species is thought to ensure they are harmless (see more in section 4.3.3 below).

There are actually plenty of variations on this theme, represented by one example in Fig. 3D. Here the studied monolayer is formed on a metallic wire,[93,270,271] or platelet[272] which are sufficiently small to make a suspension. Using electrophoresis[93,270,271] or drop-casting[272] some of these metallic objects can bridge two pre-defined stationary metallic pads, and thus form a transport path of pad-A / SAM /metallic object / SAM / pad B. In practice however, there is a variety of evidence[98,271] that one of these double junctions is shorted and only one junction is active. Adsorbing molecules with two sticky ends between two nano-particles (cf. one in Fig. 3D) forms a larger object that can now stretch across a wider trench made by lithography.[273]

Using a bridging element (e.g., Fig. 3D) also helps limiting the contact area. However, soft 'adhesion' can also work with much larger area pre-prepared metallic films, to form a standard sandwich configuration (Fig. 3A). In this method a thin metallic film is deposited on a sacrificial substrate and then transferred over the pre-made monolayer; the peeling and / or transition often employs flotation on a high-surface tension liquid (e.g., water) but not necessarily. Originally known as 'lift-off, float-on' (LOFO),[274,275] it evolved into 'polymer-assisted LOFO' (PALO),[276,277] wedge transfer[278] and 'direct metal transfer' (DMT).[61].

Alternatively, pressing-based methods use a mechanical back support that bypasses the flotation step, and induces direct film transfer, as in nano-contact printing (nCP)[150,222] or lamination.[279] These methods normally require 'sticky end-groups' with a chemical affinity to adhere to the pressed film. Film-transfer is capable of producing permanent contacts, with yields similar to those of liquid contacts. While lamination is probably easier to apply, compared to flotation transfer, both methods suffer from ill-defined interface contaminations. The major drawback of lamination is its sensitivity to the chemical nature of terminal groups. We often use LOFO for making junctions to water-compatible molecules, such as proteins.[280]

One of the most promising soft-sandwich configurations uses spin-coating of conducting polymers,[281,282] to get record high yields and reproducibility.[58] Notice though, that similar to solid film-transfer methods, spin-coating of conductive polymers is also sensitive to the solvent that is used and to how well the monolayer is wetted. PEDOT:PSS is often used for such 'soft contacts' because it is also transparent to visible light, and therefore suitable for solar-cell applications and optical spectroscopy in general. Yet, PEDOT:PSS is prone to segregation of its components or additives. Transferring a PEDOT:PSS film by LOFO overcomes some of the above-mentioned drawbacks of interface segregation / wetting.[283] In addition, the hidden assumption that the polymer does not limit the current is probably too optimistic.[284]

There are plenty of novel, creative solutions that attempt to meet the challenge of damage-free encapsulation of a molecular layer over large areas, such as use of nano-particles[285] and of graphene sheets,[286] but it is beyond the scope of this review to cover them all.

## 4.1.4. Gating and external stimuli

So far we considered only two terminal devices, based on the assumption that molecular versatility could suffice to achieve switches or rectification, without invoking a third ('gate') contact. However, the motivation to add a third, 'gate' contact is beyond imitating transistor functionality. Gating is a key handle to interrogate the system energy levels by shifting them with respect to the Fermi levels of the source and drain contacts. Large-sandwich junctions (Fig. 3A,C) are strictly impossible for gating, because only the perimeter of the extremely wide pad is accessible to gate field lines. The top contact size must be reduced to dimensions similar to the monolayer thickness (i.e., 'channel length') to facilitate gating from the surrounding, as is done, e.g., in STM-break junction formation under electrochemical gating,[287] or other single molecule junctions,[288] and may also be possible with nm-sized object assembly (Fig. 3D). A trench configuration is the standard way to apply gate bias, as is often done in 1mJ[4,289] but also in unique monolayer configurations.[234]

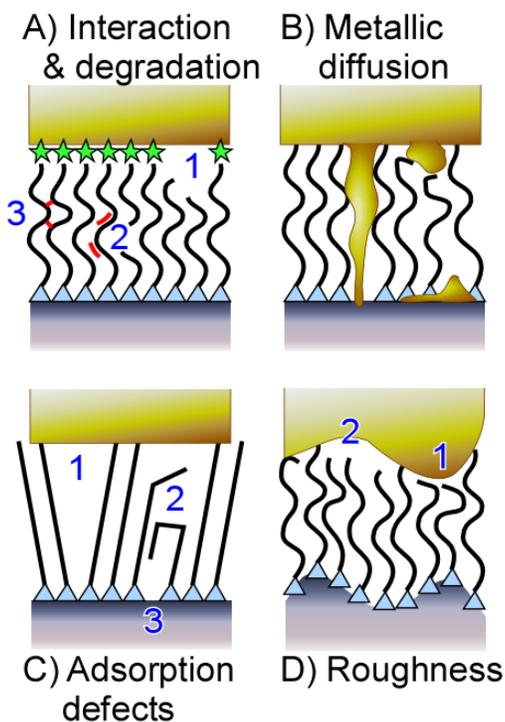

**A) Interaction & degradation**

**B) Metallic diffusion**

**C) Adsorption defects**

**D) Roughness**

**Figure 4:** Schematic description of the possible damage mechanisms and defects sources in molecular monolayers, see legends in figure.

Stars in (A) represent interaction of molecular end group with deposited metal (formation of a top chemi-contact), and digits indicate more severe damage like 1) molecular fracture; 2) formation of double-bonds; and 3) cross-linking.

Digits in (C) stand for: 1) grain-boundary in a crystalline SAM; 2) Gauche kinks and looping; and 3) missing sites.

Digits in (D) stand for: 1) a 'hot-spot' of shorter gap; and 2) a 'cold-spot' of extra

In principle, molecules have the advantage that they can be addressed not only electrically, or by light, but can also respond to pH, other chemical stimuli or mechanical deformation, which all can in principle effectively 'switch' a molecular junction. Nonetheless, it requires that the majority of molecular conductors is accessible to the external stimuli. Generally, this goal is easier the smaller is the contact (e.g., AFM[82]), but light stimuli can also be delivered to a sandwich configuration if one of the contacts is semi-transparent.[95,290]

A major open issue is the challenge of hierarchy. The main advantage of molecules is the parallel preparation of moles of identical nano-metric items. However, this also carries a severe difficulty in how to combine different items together in a pre-determined manner.[291] Attempts in this direction mostly employ scanning tips, using dip pen lithography[292] or tip-induced oxidation.[226] Examples of overcoming the next challenge of integrating a few molecular junctions into operational electronics are extremely scarce.[31,293-295]

 While this section dealt mostly with the junction's geometry and formation, the next sub-section lists a few fundamental considerations common to any junction architecture.

## 4.2. Key design considerations

There are a few concerns that are common to all large-area contacts, as schematically shown in Fig. 3. Proper attention to these aspects is probably more critical than the specific choice of assembly method.

## 4.2.1. Diffusion blocking layer

Regardless of their source, metallic atoms can diffuse through the molecular layer. Penetrated isolated atoms act as dopants,[296] or they can assemble into small clusters acting as quantum dots,[297] while metal diffusion above some percolation threshold can yield a highly conducting path shunting the desired molecules as illustrated in Fig. 4B. If such metallic filament is continuous, its conductance would be at least the quantum of conductance (77 μS) and for the whole junction, the sum over all existing filaments. This can easily exceed the combined contribution of the much larger areas surrounding the filaments, because molecular conductance of all but the shortest conjugated molecules, is many orders of magnitude smaller than the quantum of conductance.

Returning to the rationale of using Hg, it is actually contradicting. On the one hand, the success of using Hg is attributed to its liquid phase; on the other hand, Hg is clearly limited by its chemical affinity toward the bottom electrode: attempts of using a Hg drop to contact a single monolayer adsorbed on a metallic substrate, such as Au or Ag, leads to rapid amalgamation of the Hg with the underlying metal.[123] The explanation, as already summarized above, is that, given a driving force, the minute concentration of gaseous Hg, due to Hg's relatively high vapour pressure,[298] suffices to have Hg atoms cross the SAM and interact with the underlying electrode, if such reaction is chemically favourable (e.g., amalgamation of Hg with Au or Ag; the positive ΔG of the reaction provides the driving force).

At the other extreme, evaporation-induced damage is also sensitive to chemical details: the penetration of the metal through an alkyl-thiol based monolayer on a Au substrate varied with the nature of the metal, with the extreme case of evaporated Au which fully penetrated underneath the monolayer even after depositing only 10 nm Au.[223] Electrochemical deposition of Cu on monolayers of biphenyl thiols on Au was shown to form a continuous Cu layer underneath a thiol monolayer, while the monolayer remained intact.[299]

Diffusion of metallic atoms is not limited to the deposition process and can occur also from solid metals, via electro-migration. This artifact was originally mistaken as molecular switching,[300] and later evolved into an electric field-driven switching effect, known as 'memristor'.[301] LOFO-made Au contacts to alkyl-based monolayers on Si also showed memristor-like effects, which were explained as formation of metallic filaments or shorts by electro-migration in high electric fields.[302]

Diffusion across the nm-thick interfacial layer is actually much more important than is commonly assumed. It is now clear that metal penetration occurs at perfect monolayer regions due to molecular flexibility and is not limited to the presence of pre-deposition defects.[255,299] Thus the challenge of forming sustainable contacts should actually focus on how to form nm-thin diffusion barriers rather than on how to reduce the energy of the impinging species.

Non-molecular inorganic films are superior In terms of _diffusion barrier_ to organic monolayers due to their strong lateral chemical binding. In comparison, the relatively weak inter-molecular bonds in monolayers, plus their structural flexibility, often open facile diffusion routes. As a result, molecular monolayers are generally incapable of blocking the penetration of metal atoms, deposited onto them for contact formation, which can form conducting filaments between the contact and the underlying substrate (see Fig. 4B). Extended solids, such as oxides are much better in blocking such diffusion, which may explain the ability of a thin "$Ga_2O_3$ skin" on Ga-In alloy droplets (e.g., layer '3' in Fig. 3C) to form quite reproducible contacts to molecular junctions.[246]

Following this logic, Hg makes a reliable contact to monolayers on Si not because it is liquid but because Hg is one of the four elements (Tl, Bi, Pb, Hg), which are completely immiscible with Si.

Therefore, Hg has no driving force to diffuse through a SAM made on Si but a strong driving force if the SAM is on e.g., Au. Hacker and co-workers found that evaporating Au on a C18 alkyl monolayer directly bound to Si would destroy the monolayer while Ag does not. They attributed it to the different tendency of these metals to form silicides, and indeed, evaporating on an al-kyl-silane monolayer on $SiO_2$, did not cause penetration also for Au and Ti.[303] Evaporating any of the Si-immiscible elements pushes this rationale to the extreme: evaporated Pb contacts to var-ious monolayers on H-Si, yielded excellent reproducibility between repeats and replicated the results obtained with Hg.[60,304] However, evaporating Ti, which is reported to have some affinity to organic groups,[255,305,306] destroys the molecular effect, either by forming shorts (Fig. 4B) or by degrading the molecular structure (Fig. 4A).[304] The high perfection of Pb / methyl-styrene-Si in-terface is shown in Fig. 5.

There are three typical strategies to address the diffusion problem: i) increase the cohesiveness of the contacts; ii) reduce the diffusion driving force; or iii) add a diffusion barrier. They will be explained below.

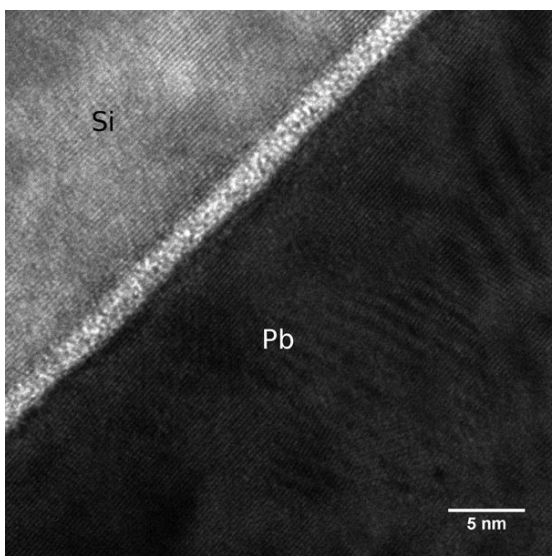

**Figure 5:** Contacts made of two im-miscible materials yield high inter-face perfection as demonstrated by a cross section TEM measurement for a Si(111)/Br-styrene/Pb junction. The spacing of the lattice fringes from Si and Pb were 0.29 nm and 0.27nm, respectively, consistent with the d-spacing of their (111) planes, reflect-ing some degree of <111> preferred orientation for the Pb. Reproduced from Ref. 304. Copyright 2013 Amer-ican Chemical Society.

<u>Contact and SAM cohesiveness</u>: While most evaporation approaches attempt extremely slow deposition, a high electro-chemical deposition rate reduces the amount of Cu penetration through a monolayer. This was attributed to fast formation of strong Cu-Cu bonds, which reduces the amount of isolated atoms that are free to diffuse.[268] In addition, the lateral attraction within the SAM can also be reinforced. Alkyl chains, even within a highly ordered, dense monolayer have sufficient motional freedom (at room temperature) to allow temporal structural deformations that provide a penetration path for the metallic atoms impinging on the top of the monolayer.[223,255] In contrast, monolayers of conjugated wires with strong lateral π-stacking are more rigid and should block such penetration for the same metals deposited under identical conditions.[223,307] Pausing the evaporation after ~0.5 nm Au deposition on thiol-terminated monolayers relaxes the interface and forms Au clusters.[257] Interestingly, these clusters behave very differently on top of a monolayer made of terthiophene-dithiol or decane-dithiol. While both monolayers show medium permeability to atomic Au (without pause), Au clusters sink into the terthiophene while being effectively repelled from the alkyl monolayer.[257] This was attributed to preferable π interactions of the thiophene rings with the Au nano-particles, but can also be attributed to the alkyl chain flexibility and to defect healing being less probable for the rigid terthiophene than for the alkyl chains. Varying the interactions of the terminal groups from van der Waals bonding (with e.g., terminal methyl groups) to hydrogen bonding (e.g., COOH) to ionic (e.g., COO⁻ K⁺) was accompanied by full penetration, partial penetration and complete blocking of evaporated Au atoms, respectively.[255] Hydrogen bonds also enhance the yield and reproducibility of molecular junctions made by object assembly.[272]

<u>Reduced diffusion driving force</u>: Terminating a monolayer with sticky-ends with chemical affinity toward evaporated atoms, such as thiols for evaporated Au, traps the evaporated atoms and reduces their tendency to penetrate into the monolayer.[152,306] General rules of the thumb sug-gest that thiol termination binds to noble metals (Au, Pt, Ag, Hg); carboxylic acid or methoxy groups bind to transition metals (Al, Cu) and methyl is non-reactive, except towards Ti or Ca, but as noted above this reaction is destructive.[305] Notice though that strong interfacial reaction (e.g., stars in Fig. 4A) might interfere with molecular properties, such as the molecule's dipole.

As noted above, adjusting the miscibility between the two contacting solids is another way to decrease the diffusion driving force. While Hg and Pb are completely immiscible with Si, evaporated Ag, which is poorly miscible with Si, was also reported to yield reliable contacts to monolayers on Si.[303] Substrates of amorphous carbon also appear stable against diffusion of a variety of evaporated metals,[158] possibly due to strong in-plane bonds.

<u>Additional diffusion barrier</u>: The lack of lateral bonds in most SAMs obviously facilitates diffusion compared to e.g., extended solids like oxides or even graphene. The presence of native oxides, either on the bottom contact (e.g., $SiO_2/Si$ or $Al_2O_3/Al$, see Fig. 3C, hatched region **1**)[86,170,303] or on the top contact (e.g., $Ga_2O_3/GaIn$, see Fig. 3C, hatched region **3**)[246,308] is well-documented to enhance the reliability of NmJ. While not metallic, such thin oxides are rarely stoichiometric and therefore their added resistance is negligible compared to that of the studied monolayer.[107,219] Nonetheless, thin oxide layers can have dramatic charging effects due to trapped charges, and significant capacitance. A foreign diffusion buffer can also be introduced deliberately (see Fig. 3C, hatched region **2**). For example graphene sheets (GS)[286,309] or ALD-deposited $Al_2O_3$[310,311] can be used to buffer between the evaporated atoms and the molecules. Graphene especially has extremely high conductivity, however attempts to use suspended graphene flakes for mm-long conductance led to impractical high sheet resistance.[286]

## 4.2.2. Surface roughness

The roughness of the stationary electrode(s), in both sandwich and trench configurations is critical for monolayer uniformity[312,313] and the resulting transport characteristics (see Fig. 4D).[70,224,314-316] Generally, the desire is to make the whole contact area as uniform as possible, to avoid a situation of 'hot-spots' of shorter transport length, and thus, if transport is by tunneling, exponentially higher conductivity. An interesting exception where 'hot-spots' are desired is that of molecular transport measurements combined with recording of a Raman signal.[92] Here a 'hot-spot' enhances the activity for both characterizations and therefore justifies their comparison. Pushing roughness to its extreme may lead to that what we conceive as an infinitely large array (NmJ), is actually a collection of a few active contacts of 1mJ junctions or of junctions of a few molecules each.

Rougher surfaces lead to higher Hg amalgamation probability[123] as well as to much poorer reproducibility[224] and even to emergence of asymmetry in current-voltage characteristics.[316] A general rule of the thumb is that the RMS of the surface roughness should not exceed 0.5 nm over 4 by 4 $\mu m^2$ area. A more detailed analysis should take into consideration also the periodicity of the roughness, since sharp protrusions are more harmful than slow changes. However we are not aware of a systematic study of this kind.

Substrate smoothness is one of the advantages of using semiconducting bottom contacts, because polished single-crystal wafers of especially Si, but also of GaAs, and GaP are commercially available, cut almost perfectly parallel to a defined crystalline plane. Surface etching, done for cleaning or patterning can induce roughness and special care has to be taken to avoid it. One of the reasons to work with the Si(111) surface rather than the industry preferred Si(100) is the superior smoothens of H-terminated Si(111), achieved by buffered (pH~8) HF treatment (e.g., $NH_4F$).[88] Also amorphous carbon (e.g., pyrolysed photoresist films)[158] naturally forms ultra-smooth surfaces.

In contrast to semiconductors, metallic substrates are used in the form of multi-crystalline thin films, with RMS roughness of ≥ 2 nm. So-called commercial 'atomic-smooth' metallic substrates or flame-annealed metallic films,[317] have ultra-smooth regions but of limited lateral dimension (~ 100 nm). While this is suitable for STM and possibly AFM studies, it does not meet require-ments for large area NmJ. There are a variety of other methods to improve the smoothness of metallic thin films, such as primer layer,[314] ultra-fast evaporation[318] or annealing[319] (see Ref. 320 for a detailed discussion on preparation of metallic substrates). The most promising method is template stripping, where a metallic thin film is first evaporated on a dummy substrate, then glued to a new mechanical support and peeled from its original substrate, to expose its 'inner' surface for monolayer formation.[321,322] If, as is usually the case, the monolayer is formed from solution, then care should be taken that the adhesive required to glue the substrate to a me-chanical support, will not be extracted during the immersion in the monomer solution and con-taminate the monolayer.

The roughness of the top contact should also be considered; for example, liquid metal top-contacts are commonly thought to be ultra-smooth due to their fluidity and tendency to mini-mize their surface energy. However, the surface oxide skin of InGa can lead to severe roughness under certain preparation procedures.[308] One of the advantages of the LOFO method is that it uses the (smoother) peeled side of the metallic film as its active surface. However, the interface is formed by repulsion of the floating solvent, a process which could wrinkle the metallic leaf and leave interfacial voids if not done properly.[275] Such voids affect not only charge transport but also energy level alignment at the interface.[323] The genuine technical difficulty in assessing the degree of roughness of buried interfaces may explain the limited answers to this critical issue.

### 4.2.3. Chemi-contacts / Physi-contacts

A common classification of molecular junctions is done according to one or two 'chemi-contacts',[28,109,239,324] where 'chemi-contact' means a specific chemical bond between the molecule and the substrate as in self-assembly (Fig. 1 middle, yellow circles).[133] A molecule with reactive groups at both ends (e.g., the blue triangle and pink circle in Fig. 3B) forms a junction with two chemi-contacts. This classification refers both to the chemical stability of the interface and to its electrical properties. The relation between chemical binding and electrical properties is less trivial than commonly assumed and will be further discussed in section 6.2. Here, we briefly touch on the synthetic requirements: how many sticky ends does a molecule need?

Within the classification of Fig. 3, sandwich junctions (Fig. 3A) generally require one chemi-contact to the substrate / bottom electrode or to the suspended object (Fig. 3D). However, sandwich junctions can be made also without specific binding groups by e.g., Langmuir-Blodgett monolayer[1] or electrostatic adsorption of large bio-molecules.[325] A trench configuration (Fig. 3B) and practically every 1mJ require two chemi-contacts. Construction of asymmetric junctions must use different contacts to ensure specific molecular orientation; such asymmetry is more difficult with a lateral trench configuration (mostly the contacts are made of the same metal) though asymmetric contacts are quite trivial for a vertical trench, and obviously with a sandwich junction.

Contact deposition procedures can also be considered in the context of universality vs. specificity. Any procedure that requires specific affinity of the top contact to the molecular terminal group (i.e., a top chemi-contact) is by definition specific. A 'universal top contact' is desired for example in the case of proteins, which do not have an identical chemistry, or for comparing contacts to a variety of metals. At this point searching for a 'universal top contact' seems too optimistic, especially acknowledging the critical role of diffusion. A variety of practical considerations, such as hydrophobicity, monolayer stability, reactivity with bottom substrate and device architecture, all affect the choice of top contact formation.

## 4.3. Do we measure genuine effects or artefacts?

For good and bad, molecules and their monolayers are soft matter, and one consequence is that monolayers are never as perfect as might appear from popular graphical illustrations ("cartoons"). Generic types of defects are schematically illustrated in Fig. 4. As detailed in the previous section, there is a lot of room for improvement by considering the nature of the molecules, the monolayer formation conditions (section 3.1) and the assembly details (section 4.2). The awareness of possible contributions by minute defects has worried the experimentalists since the days of Mann and Kuhn.[101] Similarly, their argument of 'excellent fit with theory'[101] has been used since, as indirect proof for that the data's origin is an authentic molecular effect. Here we consider in how far, for a given junction, the observables, i.e., quantities or features that can be measured /seen, are affected by defects, or, more to the point, in how far they reflect genuinely transport across the molecular junction?

### 4.3.1. General considerations

There are three main points to realise about defects:

1. Every defect counts; in other words, electronic charge transport is by far more sensitive to defects than any complementary surface analysis; because of field effects there is no clear minimal size.

2. Electrical properties often depend exponentially on molecular parameters;

3. We see the 'half-empty glass' – only defects that act to increase transport can be detected.

The first argument is statistical in nature. In contrast to 3D matter where defects such as dislocations would normally affect a macroscopic property only if they extend beyond some minimal length, or above some percolation limit, in 2D structures, one defect easily spans the '0' dimension (i.e., the thickness of the monolayer is shorter than a typical percolation length) and therefore suffices to drastically enhance the net conductance.[326] Furthermore, "correlated defects" are required to affect bulk properties, while for monolayers practically any defect counts. The probability for getting independent defects is orders of magnitude larger than to find "correlated" defects needed to affect bulk properties. This simple statistical argument is the main reason why concern over defects can never be exaggerated for NmJ.

The second reason why defects affect transport so strongly is that transport probability frequently includes exponential dependences. Examples are the exponential length attenuation of tunneling probability, exponential bias effect on transport by thermionic emission (for semiconducting contacts) and the effect of impurities on the position of energy levels and surface potential. This is different than most surface spectroscopy tools (e.g., IR, XPS), where the signal is linearly summed over the probed area. For example a change of 1 ppm or even 1 % would be hardly visible in IR or XPS spectra but could easily dominate transport. The reason is that the exponen-tial enhancement of transport by a defect can well exceed its fractional coverage, which can be no more than 10s to 100s of ppm.

Worse still, the probability to find defects crossing the full width of an insulator film reduces exponentially with film's thickness, and thus can easily be confused with exponential length-attenuation of transport by tunneling.[326] As discussed in section 6.2 below, the collective molecular electric field[51,327] has a critical effect on net transport. A defect, or a 'missing dipole' is equivalent to an opposite electric field.[328] Semiconducting contacts can help detecting defects because of their extreme sensitivity to molecularly-induced dipoles.[329] Relying on both effects we showed that flotation of a top contact (e.g., MoPALO, see section 4.1.3 above) causes no metal penetration in contrast to indirect evaporation where the photo-voltage and Schottky barrier were much smaller, similar to those without molecules.[329]

The third point to bear in mind when considering defects is that "we see only the half-empty glass". Namely, the above argument about the exponential nature of transport implies that it is sensitive to defects only if they *enhance* the current, while "dark" spots of lower current would be irrelevant as long as they cover a small fraction of the area.[323,328,330] This argument holds for inhomogeneity in either the electrostatic potential (Schottky barrier) or film thickness (tunneling barrier). While this point seems trivial it has an important consequence. It is easier to observe a molecular effect that enhances transport than the opposite. For example, a similar amount of defects would have a negligible effect on transport across a monolayer of fairly conducting molecules, but could easily dominate the transport of a monolayer of highly insulating molecules. The relative contribution of defects to net transport can change during the measurement. Thus, for example, high current fluxes through limited metallic hot spots can lead to their heating and increase their resistance.[62,331] Similarly, for an inhomogeneous Schottky barrier, bias can change the efficiency of current transport via defects, relative to that via the rest of the interface. [328,330]

Still, the resulting I-V traces can be roughly fitted to idealized (homogenous) models[326,330] (section 5) and special care is required to distinguish such artefacts-dominated transport from homogenous barrier.[62,328,330,331]

### 4.3.2. Roughness

Geometrical roughness in general is caused by either steps or poly/nano-crystallinity in the substrate (Fig. 4D) or by a non-perfect monolayer (Fig. 4C).[332] Roughness, or thickness distribution is a critical problem in trying to analyse tunneling transport data, because it affects exponentially the tunneling probability. Within the so-called Simmons model[333] (namely tunnelling across an extended insulator), simulations show that current-voltage characteristics across a defective barrier can be fitted quite adequately to models of transport across a homogenous barrier. The only difference is that defect-dominated transport appears like tunneling over a *shorter* distance of a *higher* energy barrier.[315,331] Increasing the current across the junction increases the resistance of nanoscopic filaments due to Joule heating and provides a fingerprint for defect-dominated transport.[62,331] As a result a larger portion of the high current passes through the high-resistance majority of the area region, an effect that can be quantified by elaborated models.[315] Miller et al., showed that fitting a standard 'Simmons model'[333] to a rough surface (a standard deviation in roughness of ~15% from the nominal tunnelling distance) yields a tunnelling distance which is ~33% smaller than the nominal value.[315] Considering the typical thickness of molecular junctions (~1 to 2 nm), even for an "atomically smooth" surface a roughness of ~0.5 nm presents some 50 to 25% of the insulator thickness.

Due to roughness and hot-spots, it is very plausible that charge transport is dominated by only a very small fraction ($\sim 10^{-4}$) of the nominal contact area.[19,73,257,334] This can explain some of the large variability in reported current flux per molecule.[26] Yet, there are two important cautions here. First, Selzer et al. found the currents per molecule (nitro-substituted phenylethynyl, "OPE dithiol") connected in either 1mJ or NmJ to agree well at low bias and low temperature (<0.1 V, < 50 K), while they could vary by 3 orders of magnitude at ca. 1 V and room temperature.[55] Thus, it is possible that differences attributed to roughness actually reflect genuine changes in transport characteristics. A second caution is that often the exact area is required to extract some transport characteristics, such as the energy barrier height for tunnelling,[272,335] the coupling strength,[43] or in the case of a semiconducting electrode, the Schottky barrier in the semiconductor.[323] While this dependence is often logarithmic (energy barrier $\propto$ log(area)), as the electrically active area can actually be orders of magnitude smaller than the nominal area, this can severely distort the reliability of area-dependent extracted parameters.

### 4.3.3. Reproducibility and data analysis

The direct consequence of the ultra-sensitivity of electron transport to defects is the poor reproducibility commonly encountered in molecular electronics. Combined with the log-normal distribution of transport probability[69] it often leads to orders of magnitude variations in reported current values. Reproducibility differs from 'yield': the latter refers to pre-filtering of the junctions (in order to gain 'better' reproducibility) based on some arbitrary threshold resistance for "shorted" and / or "non-contacted" (=open) junctions. This practice might have been justified in MolEl's early days, but should be discouraged now due to improved contacting processes.

Salomon et al., made an extensive survey in 2003 that showed the orders of magnitude difference between the current per molecule (for similar alkyl-thiols) reported by different labs and test-beds.[26] By 2008, the results of a similar survey by Akkerman and de Boer were already more consistent, if the nature of the contacts was taken into account (i.e., di-thiols cf. mono-thiol).[28] These statistics have been further improved since then,[19] but reproducibility, within repeating measurements *and between labs*, remain important requirements to be met to indicate that transport is indeed controlled by the chosen molecule, rather than by artefacts.

Our own experience suggests that reasonable reproducibility is a foremost indication for reliable junctions[336] more so than fitting to models, which can be very misleading.[62,326,331] A narrow distribution of data was suggested as a quality criterion also for InGa/mol/Ag junctions.[337] However the opposite is not necessarily true: tunnelling currents are exponentially sensitive to molecular details and therefore easily suffer from poor reproducibility. For example, if the source of irreproducibility is a normal distribution of monolayer thickness, then transport by tunneling will show a log-normal distribution.[69] Also properties like energy levels or dipole mostly affect the transport exponentially (via the barrier height). Therefore, averaging on log(current) is more accurate than direct (linear) averaging which gives excess weight to extrema. Yet even log-averaging may be too sensitive to values of extreme outliers and instead the median or even fitting the log(current) data to a Gaussian distribution is the more robust procedure to exclude outliers.[69]

In contrast to the current magnitude which depends exponentially on the roughness and other sources of defects, the shape of the I-V curve, or the variation of the current with the applied bias is commonly much more reproducible for repeat measurements of the same junction,[56,315,332] and therefore we strongly suggest to extend the analysis to current-voltage dependence (see section 5) and not just current magnitude.

In summary, large area molecular junctions, NmJ, can easily lead to the study of defects, thus begging the question how to minimize them so that they do not dominate the electronic transport characteristics and how to identify their contribution. No universal approach exists (yet) for preparing robust junctions or for identifying artefacts. This admittedly daunting challenge should not discourage us, because finding solutions to it is a critical enabling step toward genuine bottom-up construction of molecular electronic devices. The following two sections are dedicated to electrical performance of such large area molecule junctions.

## 5. Modeling transport across molecularly-dominated junctions

Somewhat surprisingly, the study of molecular junctions still lacks a widely accepted, simple, generic description of the current-voltage relations across molecular junctions. This section is dedicated first to why such models are needed and then bench-marking of few of the current-voltage relations which we consider most relevant. We will dwell specifically on analyses of experimental data measured for large-area Hg/molecule-Si(111) junctions, where the Si is as highly doped as possible without altering its lattice ($\geq$1E19 dopants/$cm^3$, often reported as resistivity $\leq$0.01 $\Omega$cm). Such high doping levels imply that the Si presents only a very small barrier for transport, compared to 100s of meV with medium ("normal") doped Si. Indeed, Si has a minor effect on the net magnitude of the current, but the fact that it has an energy gap affects the shape of I-V traces. For simplicity, we shall consider the Si contribution as a perturbation on what is basically molecular-dominated transport as occurring in pure metal-molecule-metal junctions. Si has many merits as substrate for MolEl, not just its technology relevance but also its versatile, resilient surface chemistry,[16,163,164] as well as reliable contact forming with Hg (previous section). Although Si's own contribution to transport adds a clear complication to modeling the net transport, it actually presents a challenging test-case on how to apply 'idealized' transport models to realistic junctions. As noted in previous sections, the capability to account for a variety of non-idealities is especially relevant for large-area, NmJ, junctions.

## 5.1. Why use generic models?

Historically, the success of the microelectronics industry is based on established approximate relations,[12] an approach that was found useful also for more exotic materials, such as polaronic and ionic ones. The MolEl community, however, still mostly relies on detailed quantum-mechanical treatments,[7] which is justified by the extreme sensitivity of net transport to atomic details. Yet, the fundamental shortcoming of quantum mechanical calculations is that their answers can only be as good as the accuracy of the input parameters (e.g., atomic coordinates), details which often cannot be verified experimentally. For example, the ubiquitous, 'standard' gold-thiol system, includes a large amount of variants on the exact S-Au bond formation (isolated thiol, pair or oligo-thiolates, with a Au adatom or bridging 2-3 Au atoms)[77] that vary with the adsorption conditions and type of molecular tails.[53] This makes any accurate computation very specific to the structural assumptions, an aspect that hinders bench-marking across different platforms, toward optimization of a given functionality or transport feature. While computation is vital for fundamental insight, it is often difficult to generalize from a specific case to a generic phenomenon, or to gain predictive power.

Thus, without detracting from the values of fundamental modeling, there is great power in 'black-box' relations that lead to generic 'figures of merits' which are, admittedly, blind to details. There are three possible motivations for data analysis, using a generic fitting approach:

1.  Insight and comparison with theory; for example, identifying a specific transport mechanism to allow comparison of experimental observations with computational results and allow predictions.

2.  Reproducibility and bench-marking; for example, how reproducible are results with a given junction? In how far do they differ from those obtained with other constructions?

3.  Predictive power; for example, which molecular change would make this junction more insulating or more non-linear?

The transport models that we refer to here are extremely simplified analytical relations. The major merit of analytical relations is that they can be applied to almost any experimental data, where the quality of the fitting is actually one of the observables. This generic nature is in fact in accordance with experimental current-voltage relations of molecular junctions that are often mostly featureless and can be fitted with very generic relations, using merely 2 to 3 free fitting parameters.[335,338] This section is dedicated to a generic description of the bias effect on tunneling current, while section 6 briefly touches on key issues, which are beyond this basic common ground. We start with a short justification of why we consider tunneling at all.

## 5.2. Hopping vs. Tunneling

Proposed transport mechanisms can be roughly divided into two families, that of tunneling and hopping.[8,123] Empirical attempts to fit MolEl current-voltage (I-V) traces to other generic models (field emission, space charge-limited)[253] generally failed, although some of these mechanisms may be relevant for describing transport across longer molecules. The temperature-dependence of transport is often considered as a clear criterion: if the transport's temperature dependence is described by $\sim exp(-E_a/kT)$ , with $E_a$ being the 'activation energy', then hopping dominates the transport,[257,296] such as, for example, injection at one of the electrodes.[332]

However, tunneling can also account for some temperature dependence, due to broadening of the Fermi-Dirac distribution of carriers in the electrodes with increasing temperature.[289,339] In such a case, the apparent activation energy is highly dependent on the relative alignment between the molecular level that is relevant for transport and the Fermi level, the "barrier height": the temperature effect increases as the barrier decreases (approaching resonance). This is a bit confusing because the tunneling traversal time also increases as the energy alignment approaches resonance, and, therefore, the probability for electron-nuclear coupling increases. A more realistic view is a continuous transition between the extremes of pure tunneling (negligible electronic-nuclear coupling) and pure hopping.[8] Apparent activation energies between 20 meV[289] and 220 meV[32] were ascribed to phase-coherent tunneling and not to hopping. Molecular junctions made of terthiophene-dithiol show similar or lower activation energy values (10 to 100 meV), depending on the contacting metal,[296] but were explained by hopping into induced states[296] rather than by level broadening.

The temperature effect also depends on the junction configuration. Returning to the 1mJ vs. NmJ comparison of phenylethynyl (OPE) derivatives,[55] the conductance in 1mJ was highly sensitive to both bias and temperature (activation energy up to 80 meV), compared to a much milder effect in NmJ of the same molecule.[55] While hopping may explain the temperature dependence, it is at odds with the strong bias effect (non-linear I-V traces) for 1mJ.  A plausible explanation is tunneling across a low barrier (1mJ) compared to high barrier (NmJ);[289,339] the reasons why the energy alignment can differ between 1mJ and NmJ are further considered in section 6.2 below.

A monolayer of alkyl-chains can even show an opposite temperature effect: a 30 fold increase in resistance with temperature, between 10K and room temperature.[340] This 'metallic-like' behavior was attributed to entropy-induced structural disorder in the flexible alkyl monolayers. Using Si(111) as a substrate forces slightly lower alkyl chain binding density compared to e.g., alkyl-thiols on Au, which possibly enables such structural flexibility.[340] Thus it appears that theoretical descriptions of charge transport across molecules should take into consideration structural and environmental effects,[341] an approach which is not that much pursued.

If one of the contacts is a semiconductor, a significant temperature-dependence can be observed even for 'far off'-resonance tunneling', for example an activation energy of ca. 50 to 100 meV for highly-doped Si/$\sim$1.5 nm SiO$_2$/Hg (see SI to Ref. 340). This effect is due to surface-charging of the semiconductor. A similar effect is expected for any non-metallic electrode, like an inorganic semiconductor, conducting polymer[284] or thin oxide barrier (see section 3.5), because the filling of the electrodes' DOS is expected to be more sensitive to temperature and less trivial to account for than in metals.

The weak to no temperature dependence in much of MolEl points to transport by tunneling. Another clear fingerprint of transport by tunneling is the exponential attenuation of current with distance, with an attenuation factor, $\beta \sim 1.2$ to $0.8$ Å$^{-1}$ for saturated organic molecules and significantly less (down to $0.2 - 0.1$ Å$^{-1}$) for conjugated ones;[5,19,28] Since this aspect is heavily covered in the literature it is dealt with here only briefly (section 6.3 below). We note, though that also with respect to the temperature dependence of conductance things are not completely clear, as evidence is accumulating that efficient transport can occur without temperature activation over distances that are well beyond those, considered feasible for measurable tunneling currents,[19,34,342] challenging accepted wisdom. Thus, "long-distance transport = hopping"[123] appears not universal, although it was demonstrated for specific cases.[10,34,343]

## 5.3. Polynomial approximation for the current-voltage relationship

With all above-mentioned exceptions, tunneling is the major transport mechanism for MolEl. Tunneling, as a quantum-mechanical process, cannot be generally described by a simple I-V relation. However, different approximations can produce simple, analytical I-V relations, described by the 'Simmons model',[333] 'super-exchange(-mediated tunneling)'[40] or Landauer model (see section 5.4 next). While the mathematical expressions of these models are very different, surprisingly they all yield an extremely similar graphical form of the I-V traces. Namely, a given experimental I-V trace can be fitted fairly well to different tunneling models, yet each approach yields different values for a physical parameter, like the barrier height.[111,272,344] The frustrating aspect is the lack of clear distinction in the typical I-V behavior predicted by the different tunneling models. We suggest that the similar graphical form is not accidental and reflects a genuine behavior of transport by tunneling, as further discussed in this section.

Regardless of detailed assumptions, any tunneling current is roughly a polynomial function of the applied bias:[7,272,335,345-347]

$$\mathrm{I} \approx G_{eq} \cdot V \cdot \left[ 1 + S \cdot \frac{V}{V_0} + \left( \frac{V}{V_0} \right)^2 \right] \tag{1}$$

where $G_{eq}$, $V_0$ and $S$, are empirical fitting parameters, named the equilibrium conductance, scaling voltage and asymmetry factor, respectively.

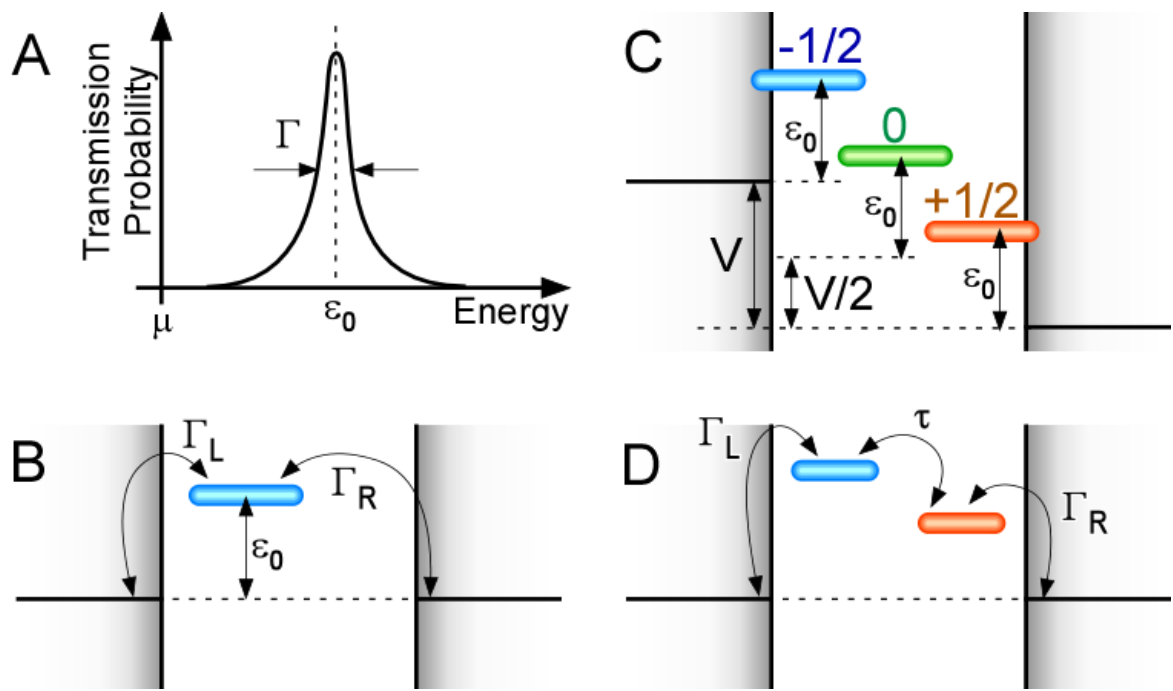

**Figure 6:** Schematic illustration of a generic single Lorentzian transport model (**A**) and few possible corresponding energy level diagrams (**B-D**). Notice that in (**A**) energy is shown on the horizontal axis while B-D show energy on the vertical axis as a function of distance across the junction (horizontal axis).

**A)** Transmission probability is commonly estimated by a Lorentzian located at a distance $\varepsilon_0$ from the electrochemical potential of the electron in the molecule ($\mu$).

**B)** A one-electron energy diagram commonly associates the transmission function (**A**) with a specific frontier molecular orbital (blue bar) at an energy $\varepsilon_0$ above (or below) the electrochemical potential of the molecule, which at zero bias roughly equals the Fermi level, $E_F$, of the electrodes (shaded rectangles on left and right). The Lorentzian width ($\Gamma$ in **A**) reflects the coupling strength between the frontier orbital (blue bar) and the two electrodes ($\Gamma_L$, $\Gamma_R$). The two coupling terms are generally different due to e.g., spatial location of the frontier level or overlapping of the molecule's valence orbitals (e.g., s, p, d) with those of the contact.

**C)** Different possible scenarios under applied bias, V. In the symmetric case, the molecule's electrochemical potential, $\mu$ is fixed at the middle of the bias window ($E_F$ -V/2), which shifts the relative position of the frontier level (LUMO, green bar) to $LUMO - E_{F,L} = \varepsilon_0 - eV/2$. If the molecule is 'pinned' to the left electrode (blue bar), $LUMO - E_{F,L} = \varepsilon_0$, regardless of the applied bias; if the molecule is pinned to the right electrode (orange bar) $LUMO - E_{F,L} = \varepsilon_0 - eV$. Colored numbers above each bar give the corresponding asymmetry, $\alpha$ (see text).

**D)** A maximum in the transmission function can arise also from multiple, spatially localized molecular levels; in such a case, the identification of the transmission parameters ($\varepsilon_0$, $\Gamma$, $\tau$) with the molecular density of states is non-trivial.

Differentiating the current with respect to voltage yields the conductance, $G$:

$$\frac{dI}{dV} = G(V) \approx G_{eq} \cdot \left[ 1 + 2S \cdot \frac{V}{V_0} + 3\left(\frac{V}{V_0}\right)^2 \right] \qquad (2)$$

The polynomial coefficients are based on the Taylor expansion, and therefore can be computed from the value at 0 V of the first ($G'$) and second ($G''$) derivatives of the conductance.

The equilibrium conductance is the conductance at 0 V (in physics this is often simply called *the* conductance), measured in $\Omega^{-1}$:

$$G_{eq} \equiv G(0V) \propto N \cdot e^{-\beta L} \qquad (3)$$

The equilibrium conductance is linearly proportional to the number of molecules, $N$ (i.e., parallel conductors) or active area, and exponentially inversely proportional to the tunneling distance, $L$ (or molecular length).

The scaling voltage, in V, expresses the characteristic voltage scale:

$$V_0 \equiv \sqrt{6 \cdot G(0V)/G''(0V)} \propto \varepsilon_0 \qquad (4)$$

Now, if one takes any tunneling based I-V trace and plot it as $I/G_{eq}$ against $V/V_0$, all traces will roughly fall on a single generic trace.[272,338,347]

The scaling voltage, $V_0$ is close to another commonly used parameter, known as transition voltage, $V_t$ (see section 5.5 below). A different terminology of 'critical'[338] rather than 'scaling' parameters has also been suggested, though it is our view that 'scaling' suits better the physical meaning of this parameter (in analogy to characteristic length scales or energies) while 'critical' hints to phase transitions, which is not relevant for tunneling. In terms of fundamental tunneling, $V_0$ is closely related to the injection barrier (or tunneling barrier height), $\varepsilon_0$, which is identi-fied with the energy difference between the Fermi level and the nearest molecular orbital (see Figs. 2 and 6B). A complementary determination of the molecular energy levels can be done experimentally by different variations of photo-electron spectroscopies (PES)[8,43,113,197,211,348] or computationally by DFT. However, deposition of, or even mere contact with a metal is known to affect the energy alignment within the molecule (see section 6.2 below) and, thus, it is unclear in how far *surface* characterizations like PES, are relevant to the energetics of a molecule contacted from *both sides*.

The 3[rd] parameter is the asymmetry (dimensionless), which is expected to be rather small ($\rightarrow$0):

$$S \equiv G'(0V) \Big/ \sqrt{\tfrac{2}{3} \cdot G(0V) \cdot G''(0V)} \qquad (5)$$

Graphically, Eq. 2 describes a parabola with a minimum at: $V_{min} = -S \cdot V_0/3$. However, back and forth tunneling probabilities are basically identical; tunneling which is asymmetric with respect to bias polarity is possible only in cases of a significantly asymmetric potential profile across the junction or when there is a forbidden gap in one of the contacts (e.g., semiconducting or magnetic contacts). Thus, S and any higher odd terms in the polynomial expansion of the expression for conductance (Eq. 2) are fundamentally smaller than the even ones.

The adequacy of the Taylor expansion (Eq. 2) to describe experimental data is tested in Fig. 7A, which plots the conductance (by numerical differentiation[349]) of 5 different Hg/monolayer-Si junctions. The junctions were generally with highest possible Si doping, either p- (red, magenta) or n-(black, blue), with two lengths of saturated alkyl chains (C16: black, red; C18: blue, magenta). The fifth set shows transport across Hg/Br-styrene-Si junctions with moderately-doped Si. This junction shows Ohmic behavior,[350] with almost no barrier within the Si[46,85] and this could be interpreted as tunneling-controlled transport. However a quick look at the *G-V* plot (Fig. 7A) reveals that this junction does not have parabolic *G-V* behavior, and therefore is readily identified as non-tunneling dominated.

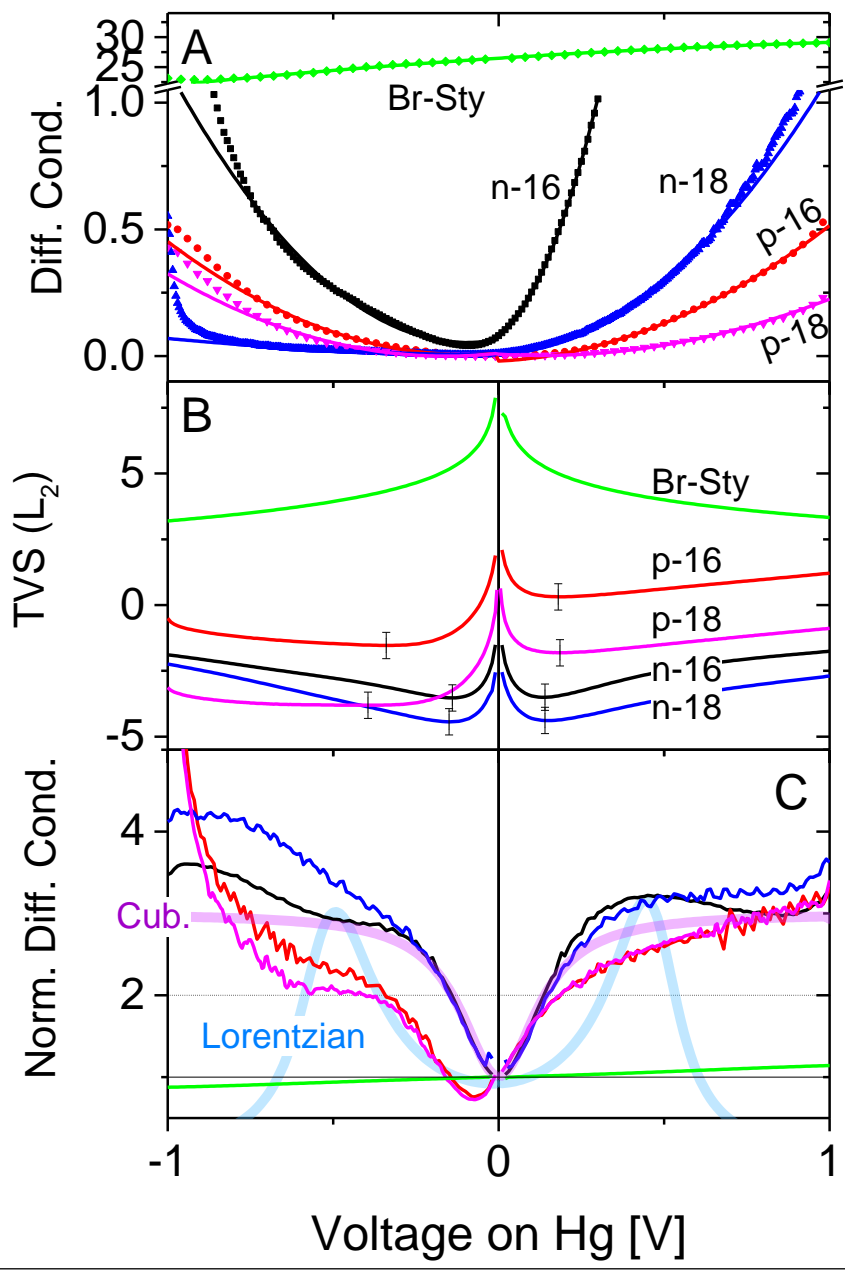

**Figure 7**: Three alternatives for empirical analysis of tunneling current-voltage curves, showing:

   A. Differential conductance;
   B. Transition voltage spectroscopy (TVS) plots of the $L_2$ functional (Eq. 10);
   C. Normalized differential conductance (NDC, Eq. 11).

An identical set of raw-data is used for all three panels, showing two lengths of alkyl-chains (C16, C18) on highly doped n- or p-Si, as well the Br-styrene monolayer on moderately-doped n-Si. See legend in figure. In all cases the top contact is Hg; Solid lines in **A** are polynomial fits (Eq. 2); vertical black lines in **B** mark the minima points ($V_t$); pale, thick lines in **C** are cubic approximation (purple, Eq. 2) and a single Lorentzian (blue, Eq. 6), and horizontal dashed line marks NDC=2. Original data are taken from Refs. 46, 340, 350.

Solid lines in Fig. 7A are fits by the approximated polynomial relation (Eq. 2). As can be seen, the fitted curves diverge from the raw data as the bias is increased from 0 V. Often, asymmetric relations do not fit well to Eq. 2, and in practice, separate fits for positive and negative bias polarity are made. The main merits of the generic polynomial approach of Eq. 2 are:

- No *a priori* assumptions;

- Effective separation between the equilibrium tunneling probability ($G_{eq}$) and its sensitivity to bias ($V_0$).

- Generic mathematical relations (conductance is parabolic with bias) that can reasonably describe vastly different cases, at least over a limited range.

A reasonable fitting of metal/molecule/semiconductor data (Fig. 7A) to the generic polynomial relation (Eq. 2) is not trivial, because we know that the barrier posed by the Si space charge does play a role (as is evident from e.g., the pronounced effect of doping type on the shape of the *G-V* curves in Fig. 7A). Still we can apply the generic parabolic G-V relation (Eq. 2) for markedly different tunneling junctions.

Actually, the observation that a very simple (and simplistic) relation like Eq. 2 (see next section for an alternative model) can be used to fit data for vastly different molecular junctions, is not as surprising as it may seem.[338] The parabolic bias-dependence of tunneling was recognized nearly half a century ago[345] and is often quoted as one of the 'Rowell criteria' to recognize tunneling.[351,352] Thus, tunneling via molecules follows the same fundamental physics of that via inorganic insulators – nature is universal, after all. Later, Holmlin et al., noted that the 'Simmons model'[333]) and the molecular view of super-exchange-mediated tunneling[40] can be mapped onto each other.[344] The major merit of the Taylor-expansion approach (Eqs. 1-5) is that it provides a mathematical translation (Eqs. 3-5) between generic G-V behavior (Eq. 2) and basically any detailed tunneling I-V model.

This is not to say that details are unimportant; the challenge is in identifying them. The parabolic relation (Eq. 2) is important in this sense, because it helps to distinguish the common standard from additional specific features. A parabolic G-V, or cubic I-V curve is the correct fingerprint of tunneling, rather than linear I-V (correct only for $V \ll V_0$), or 'sinusoidal', 'S-shape' and other creative descriptions. It differs significantly from a diode behavior ($I \propto \exp(V)$), and actually, McCrery and Bergren pointed out that the generic $\propto V^3$ shape is analogous to two diodes connected back to back, which has direct implications for electrical circuits.[31] The quality of the fitting between experimental current-voltage traces and a specific model, can be used as a proof for the relevance of a given transport mechanism only if the adequacy of the fit is significantly better than that of a generic polynomial fit (e.g., extends over wider bias range).

Adding more (>3) fitting parameters, as is often the case in detailed tunneling relations (e.g., 'Simmons',[333] 'super-exchange'[40] or Eq. 6 below) practically leads to non-converging optimization, with few possible solutions (i.e., different sets of fitting parameters yield same fit-quality).[111,335]

The scaled nature of Eq. 2 implies that $G_{eq}$ and $V_0$ are basically independent of each other. This is specifically useful regarding the large uncertainty in electrically active contact area, or the number of molecules that are actually involved in the transport in any but true single molecule (e.g., STM, break junction) experiments (see section 4.3 above).[73,335] The current scaling, $G_{eq}$, is linearly proportional to the active area (see Eq. 3), while the bias scaling, $V_0$, is independent of it (Eq. 4). Such clear elimination of the unknown contact area is often impossible in more accurate / detailed models.[272,335]

While specific tunneling models cannot generally be identified based on the adequacy of fit, they do differ in how the extracted parameters vary with controlled parameters,[272,347] such as the dependence of the scaling voltage ($V_0$ or $V_t$) on the molecular length. Thus, 'strong coupling' (applicable for extremely short distances or highly conjugated molecules) predicts the scaling voltage to be independent of the tunneling distance, $V_0 \propto \varepsilon_0$.[353] Yet, other scenarios predict more complicated relations, such as: $V_0 \propto \sqrt{\varepsilon_0}/L$ or $V_0 \propto \varepsilon_0/n$, where $L$ or $n$ are the molecular length (Simmons) or the number of super-exchange steps,[347,354] respectively. Such reciprocal dependence of $V_0$ on $L$ was reported for junctions made with saturated alkyl chains,[347] and is not necessarily an indication for tunneling through space.[353] Similarly, the way the barrier height, $\varepsilon_0$, affects $G_{eq}$ and $V_0$, varies between different models, and, therefore, a search for internal correlations between them[272] may help to support specific tunneling scenarios. Reasonable correlation between experimentally extracted and theoretical, computed values is another way to bolster the relevance of a specific model.[43,354,355] However, too often molecular electronics computations relay on *a priori* assumptions, which are impossible to verify experimentally,[53] which then can make this a somewhat circular endeavor.

## 5.4. A single Lorentzian transmission function

The empirical nature of the parabolic relation (Eq. 2) is a merit as a bench-marking platform, but clearly a disadvantage when insight into the underlying transport physics is sought.[354] While many of the initial works in molecular electronics used the Simmons model,[101,253,344] its adequacy for describing molecular junctions was criticized because of internal inconsistencies (e.g., fitting current-length or current-voltage yields different parameters) and incapability to account for molecular features (e.g., conjugated vs. saturated chains). Instead, the Landauer scattering for-malism is now widely accepted.[7,339,346,356-359] Somewhat confusingly, several closely related models such as (off-)resonant, Breit–Wigner[7,346,356] or Newns-Anderson,[359] yield an almost identical current-voltage relation. To avoid those minor differences between these models, we will describe the model as 'single Lorentzian', reflecting the shape of the transmission function. The adequacy of this model to large-area molecular transport (NmJ) can be questioned: NmJ are 2D structures, while scattering requires 1D constrictions (see section 2.2), namely localization of the passing electrons into individual molecules within the laterally infinite monolayer. In addition, the relation between the transmission function and the molecular details, is somewhat obscured in detailed computations and not intuitively clear (though we suggest some crude guidelines in section 6).

The single Lorentzian description starts from a generic Landauer formalism, and simplifies it by two drastic approximations:

- The transmission function is approximated as a single Lorentzian-shaped peak ($T(E) = 4{\Gamma_g}^2/[(E - \varepsilon_0)^2 + {\Gamma_a}^2]$), centered at $\varepsilon_0$ w.r.t. the "mid-gap" electron energy of the molecule ((LUMO + HOMO)/2, i.e., Mulliken electronegativity) in the junction, and with a full width at half maximum of $\Gamma$ (see Fig. 6A);
- The Fermi-Dirac occupation of the density of states on the electrodes is replaced by a delta-function (i.e., 0 K), which allows for analytical solution.

This yields the following analytical relation for the conductance:[360]

$$G_{(0K)} = NG_0 4\Gamma_g^2 \left\{ \frac{1/2 + \alpha}{[\varepsilon_0 + (1/2 + \alpha)eV]^2 + 4\Gamma_a^2} + \frac{1/2 - \alpha}{[\varepsilon_0 - (1/2 - \alpha)eV]^2 + 4\Gamma_a^2} \right\} \qquad (6)$$

where $G_0$ is the quantum of conductance, $N$ is the number of molecules conducting in parallel and $\varepsilon_0$ is the center of the peak in the transmission function (see Fig. 6A). $\Gamma_g$ and $\Gamma_a$ stand for the geometric and arithmetic averages, respectively, of the coupling strength to the left and right contacts ($\Gamma_L$, $\Gamma_R$ in Fig. 6C). The width of the transmission function ($\Gamma$ in Fig. 6A) in principle reflects the level broadening observed by UPS (left side of Fig. 2), though DOS and transmission are not identical properties. Finally, $\alpha = 0 \pm \frac{1}{2}$ is an asymmetry factor (equivalent to $S$ in Eqs. 1, 2, 5). This parameter measures how much the bias partition differs from 1/2 ($\alpha=0$, symmetric case), as illustrated in Fig. 6C, which is an overlay of three alternative scenarios. Each colored bar represents the same molecular level that is situated at $\varepsilon_0$ above the Fermi level at zero voltage. Applying a bias, V shifts the relative position of the molecular level. In the symmetric case, ($\alpha=0$, green bar) the level is at a constant distance from the middle of the bias window. For the extreme case of pinning to the left contact ($\alpha=-1/2$, blue bar), the position of the level is fixed with respect to the left Fermi level, while the right contact scans across it. The orange bar is the opposite extreme of pinning to the right contact. Bias partition is further discussed below (section 6.1).

Eq. 6 can be integrated with respect to bias, to get an exact analytical expression for the current:[357,359]

$$I_{(0K)} = \int_0^V dV \cdot G_{(0K)} = NG_0 \cdot 2\frac{\Gamma_g{}^2}{e\Gamma_a} \cdot \left[ tan^{-1}\left(\frac{\varepsilon_0 + (1/2+\alpha)eV}{2\Gamma_a}\right) - tan^{-1}\left(\frac{\varepsilon_0 - (1/2-\alpha)eV}{2\Gamma_a}\right) \right] \quad (7)$$

When the arguments of the inverse tangents in Eq. 7 are large, their difference can be *approximated* as:[359]

$$I_{(0K,\varepsilon_0 \gg \Gamma_a)} \cong NG_0 \cdot 4\Gamma_g{}^2 \cdot \frac{V}{(\varepsilon_0 + \alpha eV)^2 - \left(\frac{eV}{2}\right)^2 + 4\Gamma_a{}^2} \quad (8)$$

A modification of Eq. 8 was suggested in Ref. 338, which was shown to describe I-V characteristics of a large number of molecular junctions. We note, though, that the simple approximation of Eq. 1 can also be fitted successfully to at least as wide a range of different types of molecular junctions[335] (fits to both eqs. 1 and 8 are shown in Fig. 8A). Therefore, the question remains how to choose between these two presentations. Here, rearranging Eq. 8 does yield a _unique_ finger-print of the single Lorentzian model:

$$\frac{V}{I} \cong \frac{1}{NG_0 \cdot 4\Gamma_g{}^2} \cdot \left[ (\varepsilon_0 + \alpha eV)^2 - \left(\frac{eV}{2}\right)^2 + 4\Gamma_a{}^2 \right] \qquad (9)$$

Explicitly, the single Lorentzian model predicts that plotting of V/I against V (i.e., R-V) yields an inverse parabola with a maximum near (0,0). Such a comparison is made in Fig. 8B, for a junction of HD n-Si-C16/Hg. As can be seen, for this specific junction the parabolic expansion of the conductance (Eq. 2, Fig. 8B) describes the transport more adequately than the single Lorentzian (where the resistance – voltage is an inverse parabola). We argue that direct fitting to the raw I-V curves (Fig. 8A) is not sufficient to identify specific transport mechanism. Instead, higher order analyses, such as dI/dV or V/I and their variation with the applied bias, should be used to judge the relevance of specific tunneling mechanisms.

While a single-Lorentzian I-V relation (Eq. 7) is probably the best analytical description at hand, it is important to be aware of its limitations. First, neglecting temperature in Eqs. 6-8 ignores the Fermi-Dirac broadening of the contacts' DOS. This is often acceptable, as the Fermi level broadening starts to be important only when the molecular level approaches resonance with the contact's Fermi level (e.g., at $eV=\varepsilon_0/\alpha$). In practice such resonances are rarely observed, probably because the junction is anyway composed of a distribution (in time as well as spatially for NmJ, see left side of Fig. 2) of molecular energy levels (though see section 5.2 above for cases where temperature is important).

The choice of a Lorentzian-shaped transmission function is not obvious; another option is the Gaussian-shape peak;[361] such broadening better describes a statistical ensemble of different molecular configurations. The concept of single peak transmission function is often confused with 'single molecular level'. The transmission function is the quantum-mechanical output of the integral junction, which in principle can be composed of a series of super-exchange steps along localized states (e.g., Fig. 6D). While pioneering attempts to explain rectification have used a two-level transmission function,[357] a recent DFT work yielded a transmission function largely dominated by a single Lorentzian (near $E_F$) even for a strongly decoupled donor-bridge-acceptor molecule.[362] As the bias increases, further levels are expected to contribute to the transport,[346] and even the transmission function itself changes at high bias.[362] A single Lorentzian was also reported to adequately describe rectifying junctions where the frontier orbital is localized on a central ferrocene unit,[339] indicating that this approach has broader application than merely 'single level' as in fully conjugated molecular wires.

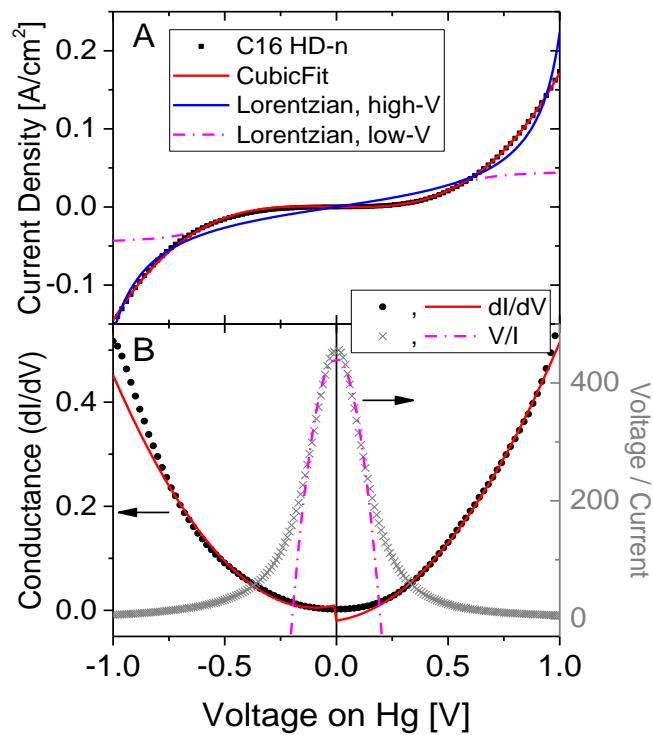

**Figure 8**: Comparing cubic approximation to single Lorentzian for fitting of I vs. V (**A**), G vs. V (**B**, left Y) and V/I vs. V (**B**, right Y) traces. Raw data (dots) are for C16 alkyl on highly-doped n-Si. Red line is cubic fit (Eqs. 1, 2 for A and B, respectively). Two fits to a single Lorentzian are shown: solid blue line is a direct fit of the current to Eq. 8 over the full bias range; dashed-magenta is a fit of V/I to Eq. 9 between ±0.15 V. Raw data are identical to black trace in Fig. 7 and taken from Ref. 46.

Similarly, the width of the transmission function relates to the coupling strength from the contact into the level and out of the level (see Fig. 6B) and does not reflect the broadening of the molecular level. The major direct link between the transmission function and molecular identity is the energetic position of the transmission function, which is identified with the energy of the level that dominates the transport (both noted by $\varepsilon_0$ here, Fig. 6B). Although the linkage to 'intrinsic' molecular properties is a bit naïve (see also section 6.2 below) the single Lorentzian approach is an efficient framework for comparing experimentally extracted values ($\varepsilon_0$, $\Gamma$) with a full quantum-mechanical computation of the transmission function.[43,346]

To summarize, the R-V approach (Eq. 9) provides a unique fingerprint for the single Lorentzian model (cf. raw I-V fitting), which is yet to be tested on experimental curves. Both Eq. 2 and Eq. 9 are parabolic with voltage but for roughly inverted quantities (dI/dV or V/I, respectively) and therefore their adequacy in describing a given experimental I-V trace should be distinctly different. Yet, for comparative purposes, the parameters extracted by these two methods can be used. Casting Eq. 6 into the Taylor-coefficients (Eqs. 3-5) yields: $G_{eq,L} = 4NG_0 \left(\frac{\Gamma_g}{\varepsilon_0}\right)^2 / (1 + r)$;

$V_{0,L} = 2\varepsilon_0 \sqrt{\frac{1+r}{(1-r/3)(1+12\alpha^2)}}$ and $S_L = 4\alpha / \sqrt{(1 - r/3)(1 + 12\alpha^2)}$, where $r = \left(2\frac{\Gamma_a}{\varepsilon_0}\right)^2$ and both

r and α $\ll 1$, so that $G_{eq,L} \cong 4NG_0 \left(\frac{\Gamma_g}{\varepsilon_0}\right)^2$; $V_{0,L} \cong 2\varepsilon_0$ and $S_L \cong 4\alpha$. Notice that the exact 'single Lorentzian' (Eqs. 6-7) offers 3 fitting parameters ($N$, $\Gamma_g$ and $\Gamma_a$) where the parabolic approxima-tion (Eq. 2) has only one parameter ($G_{eq}$). However, in the majority of cases, $\Gamma_a$ is negligible with respect to $\varepsilon_0$ ($r \ll 1$) and therefore it is practically impossible (at least for low-coupling cases, i.e. $\Gamma \ll \varepsilon_0$) to independently extract both $N$ and $\Gamma$, let alone to distinguish between $\Gamma_g$ and $\Gamma_a$ (i.e., between the coupling to the left and right contacts). The next section is concerned with practical methods for extracting characteristic parameters even without any fitting.

## 5.5. Transition Voltage Spectroscopy and extraction of characteristic parameters without fitting

As demonstrated in Fig. 8, simplified models rarely show an ideal fit to experimental data. This raises a variety of technical issues, such as which model to use (e.g., Eq. 2 or 9)? The choice of range of data to use for fitting is important, because the fitting range does alter significantly the extracted parameters;[253,335,363] there is also the fact that plotting differential quantities (dI/dV or V/I) increases noise considerably. Because Eq. 2 is an approximated Taylor expansion, its accuracy improves as the fitting is limited to the vicinity of 0 V, ca. |V|≤$V_0$/10. However, practical measurement (signal to noise) considerations set a lower limit to currents that can be measured and to the minimal step of the voltage source. In addition, transport by tunneling is often claimed to show a 'zero-bias anomaly', especially at low T,[257] which can decrease the relevance of such 'near-equilibrium' measurement approach.

The first parameter of current scaling, $G_{eq}$, can actually be extracted in a straightforward manner from the slope of the *I-V* trace near 0 V, and does not require an extended fit. As noted in Eq. 3, this parameter contains the length attenuation of the tunneling, for almost any tunneling model, and has been long recognized as a valid 'figure of merit' of molecular junctions.[28,111,296,364]

The difficulty starts with an accepted way to quantify the effect of the bias on the transport or how 'non-linear' are the current-voltage curves, which we term bias-scaling, $V_0$ and others prefer to directly identify with the barrier height, $\varepsilon_0$.[43,353,355,359,361,365] While fitting is an option, it is prone to large uncertainties, as explained above. A more deterministic, robust approach to extract $V_0$ is by the so-called transition voltage spectroscopy (TVS).[56,355] TVS is truly a powerful experimental handle, because of its simplicity and the correlation found between transition voltage ($V_t$) extracted from $I\text{-}V$ traces and molecular levels extracted from UPS[43,56,355], correlations that are remarkable as UPS missed any effect of the top contact on the energy alignment (see section 6.2 below). It is our understanding that the transition voltage, $V_t$, and scaling voltage, $V_0$, are very closely related properties.[347] While these two approaches often yield somewhat different values[363] there is no objective judgement which approach is more authentic or reliable as further explained in this section.

The $V_t$ concept is confusing because it was born by mistake. Originally, $V_t$ was attributed to a transition between simple tunneling and field-emission tunneling (Fowler-Nordheim).[56] It is borrowed from inorganic tunnel barriers where the applied bias is distributed along the tunneling gap, yielding a 'trapezoidal-shape' potential profile (normal tunneling), that turns into a triangular shape (field-emission tunneling) when the applied bias is larger than the tunneling barrier.[333] In contrast, recent views of molecular junctions assume that hardly any of the potential drops over the molecule, but falls mostly on the contacts (see Fig. 6C). Thus, likely most of the molecular junctions do not reach field-emission under any circumstances. Still, the mathematical formalism turned out to be very useful. The common Fowler-Nordheim presentation[272,366] plots $L_2$ against 1/V, where $L_2$ is defined as:

$$L_2 = \ln \left| \frac{I}{V^2} \right| \tag{10}$$

The transition voltage, $V_t$ is simply the *mathematical* minimum of the $L_2$ function and *not a change* in transport mechanism.[347] This is demonstrated in Fig. 7B for the same raw I-V data used for Fig. 7A. The vertical black bars indicate the minima on each curve. Because of the (wrong) Fowler-Nordheim origin of the TVS analysis, it often uses 1/V as its independent, horizontal axis.[56] However, the mathematical minimum of $L_2$ (Eq. 10) is the same with respect to either V or 1/V.[338,347] A further evidence that Fowler-Nordheim tunneling is irrelevant for most of MolEl is that molecular current-voltage curves hardly show a linear dependence of $L_2$ on 1/V.[272,366] For this reason, Fig. 7B uses the direct bias as its independent axis.

Conceptually, the merit and flaw of TVS is contained within the term 'spectroscopy'. On the one hand, $V_t$ values are often highly reproducible even in cases where the net current magnitude varies by orders of magnitude.[56] This feature is, seemingly, spectroscopic – a transition which is insensitive to the net signal magnitude. Spectroscopy reflects internal modes or energy levels. However, the transport mechanism is perfectly continuous around $V_t$: neither the transport mechanism is changing nor any electronic or vibronic mode is excited. $V_t$ (like $V_0$) is really an empirical measure of how sensitive is the junction's conductance to the applied bias.

Elucidating the relation between $V_t$ and the energy levels of the molecules can be done by casting the empirical $L_2$ functional on specific current-voltage relations.[347] The mathematical minimum of $L_2$ is identical to requiring:

$$\frac{dI}{dV} \cdot \frac{V}{I} \bigg|_{V_t} \equiv 2 \qquad (11)$$

The value '2' reflects the power of $V^2$ in $L_2$ and a general $L_q = \ln(I/V^q)$ with an arbitrary power $q$ ($q>1$) has a minimum at $dI/dV \cdot V/I = q$. Eq. 11 implies that $V_t$ marks the bias where the I-V dependence changes from sub-quadratic to super-quadratic.[113,347,367] Namely, the transition volt-age, $V_t$ (minimum of $L_2$, Eq. 10) and the scaling voltage, $V_0$ extracted from a polynomial fit (Eq. 2) are two manifestations of a similar property. Actually, if the 3$^{rd}$ order polynomial approximation is substituted into Eq. 11 (Eq. 1 and 2 for $I$ and d$I$/d$V$, respectively), it yields an identity: $V_t = V_0$. Yet, the formal definition of $V_0$ is as close as possible to 0 V (Eq. 4), and therefore it is obvious that a low-bias ($V_0$) and moderate bias ($V_t$) do not yield identical results.

Unfortunately in the literature there is often no distinction between the *technical tool* called, $V_t$ and its *interpretation* within the single-Lorentzian model, described in the previous section. As stressed above, not all MolEl can be described by a single Lorentzian (e.g., Fig. 8), but almost any I-V trace can be analyzed using the $V_t$ tool.[347] Still, because $V_t$ is often related to a single Lorentzian, we shall use this model to exemplify how $V_t$ can be translated into fundamental transport parameters.

Inserting Eqs. 6 and 8 into Eq. 11 yields:[353,359,361,365]

$$V_t = \frac{\varepsilon_0}{-2\alpha \pm \sqrt{\alpha^2 + \frac{3}{4}}} \xrightarrow{\alpha=0} \pm \frac{2}{\sqrt{3}} \varepsilon_0 \qquad (12)$$

In comparison, the exact definition of $V_0$ yields, $V_{0,L} \simeq 2\varepsilon_0$ (see end of former sub-section), which, thus, leaves a factor of 1.7 between $V_0$ and $V_t$, within the "single Lorentzian" view.

The two methods are compared for Hg/alkyl-Si junctions in Fig. 7A (parabolic G-V) and 4B (TVS plot), and the extracted $V_t$ (by TVS) and $V_0$ values are compared in Table 1.

Any data analysis has both a qualitative and quantitative purpose. It is clear that the TVS presentation skews the data and does not provide a qualitative indication of what is the mechanism that produces this trace. For example, the "abnormal" Br-styrene junction, specifically, looks almost normal in the TVS presentation,(Fig. 7B) while the *G-V* plot (Fig. 7A) shows immediately that this is not a tunneling trace. Despite the qualitatively correct trend, the polynomial fits (solid lines in Fig. 7A) could not be extended over the full bias range and were fitted for the positive and negative ranges, separately. Often, the polynomial approximation cannot even cover the range from 0 to ±1V, and the fitting range was arbitrarily truncated to get a good fit over the lower bias region. These are certainly disadvantages of fitting-based procedures. In terms of quantitative values, Table 1 shows that for p-Si junctions there is a good correlation between $V_t$ and $V_0$ but for n-Si, $V_0$ is about twice $V_t$. As explained above, this inconsistency reflects the sensitivity of fitting to the selected bias range for fitting. Standard fitting procedures are naturally inclined toward high signals that have a larger contribution to the averaged mean-square error. Some randomness in human choice of fitting range combined with fitting being more time-consuming compared to single point extraction explains the major drawbacks of the G-V parabolic fitting approach.

On the other hand, the major drawbacks of the $V_t$ approach are its lack of continuous, qualitative information, and the requirement for actually reaching the specific $V_t$ bias (ca. 0.5 to 1 V), which could be a real limitation for sensitive / unstable junctions. A third method, that apparently combines the merits of both former methods, is that of 'normalized differential conduct-ance' (NDC),[368] which is the common name for the mathematical term on the left side of Eq. 11. NDC is popular in scanning tunneling spectroscopy (STS) for detecting the density of states of the surface species.[368] Thus NDC is also used as a spectroscopy tool, revealing the energy levels of the *surface*. A real 'spectroscopic' consideration of peak position in NDC requires direct conductance recording rather than numerical differentiation. Still, we suggest that NDC can be adopted for examination of trends and threshold values, rather than accurate peaks. To this end NDC can be numerically computed by differentiating the log of the current with respect to the log of the bias ($NDC \equiv \frac{dI}{dV} \cdot \frac{V}{I} \equiv \frac{d \ln I}{d \ln V}$).

**Table 1**: Parameter extraction by different methods (see text for explanation of abbreviations) applied to the I-V curves of Fig. 7.

| Molecule | Si | Range | $V_t$ | $V_0$ | NDC=2 |
|---|---|---|---|---|---|
| C16 | p-HD | V < 0 | -0.34 | 0.39 | -0.34 |
| | n-HD | | -0.14 | 0.24 | -0.14 |
| C18 | p-HD | | -0.40 | 0.49 | -0.45 |
| | n-HD | | -0.15 | 0.27 | -0.15 |
| C16 | p-HD | V > 0 | 0.18 | 0.19 | 0.17 |
| | n-HD | | 0.14 | 0.35 | 0.14 |
| C18 | p-HD | | 0.19 | 0.20 | 0.19 |
| | n-HD | | 0.14 | 0.28 | 0.15 |
| **Br-Sty** | n-MD | | -- | 9.1 | -- |

It follows that the NDC is a rough measure for the power dependence of the current-voltage relations. Namely, if $I \propto V^q$, than NDC$\rightarrow q$ at high $V$, while the proportionality factor is eliminated by its built-in normalization (i.e., multiplication by V/I). Therefore, *I-V* curves which originally differ by orders of magnitude are brought into a common ('universal') NDC scale, with a lower limit of NDC$\equiv$1 for strictly Ohmic relations ($q$=1 or $I \approx G_{eq} \cdot V$, marked by lower horizontal line in Fig. 7C). Thus, if the Br-Styrene curve (green in Fig. 7) has a more or less constant conductance (Ohmic, see Fig. 7A) the NDC presentation brings it near unity, regardless of its much higher $G_{eq}$ compared to the other sets of Fig. 7. Similarly, the exponential length attenuation between C16 and C18 molecules (black / blue or red / magenta) largely disappears in the NDC presentation (Fig. 7C).

Thus the NDC provides a robust tool for exploring the *shape* of the I-V curve, regardless of its magnitude ($G_{eq}$) rather than distorting it to reveal an artificial minimum (as in TVS, Fig. 7B). As a comparison, the semi-transparent purple curve in Fig. 7C gives the NDC function of a parabolic-simulated *G-V* curve (Eq. 2) using $G_{eq}$=0.01 $\Omega^{-1}$cm$^{-2}$, $S$=0 and $V_0$=0.14 V (i.e., similar to the polynomial coefficients of C16 n-Si). This approximation is limited to a power of 3, and therefore the NDC curve that it yields, approaches asymptotically the value of NDC=3. The inflection occurs around $|V|>0.3V\approx2V_0$, and above it the measured data deviate considerably from the simplistic polynomial model. In addition, the pronounced dip below the theoretical lower limit of NDC $\geq$ 1 for the p-Si curves, reflects the plateau in the conductance plots for these curves and cannot be explained by the polynomial approximation; it is a characteristic of the Si forbidden energy gap as further considered in section 6.1. A 'single-Lorentzian' model (Eq. 6) yields a distinct NDC shape shown in Fig. 7C by a semi-transparent blue line. The pronounced peaks are characteristics of the resonances predicted by this model, and in principle should resemble the transmission function. As demonstrated in Fig. 8B, the Si-based junctions cannot be described by single Lorentzian, and the NDC presentation (Fig. 7C) confirms this.

Therefore, the NDC presentation tolerates various tunneling possibilities without involving *a priori* assumptions. In addition to the qualitative information, the NDC approach allows for unambiguous extraction of a characteristic bias scaling: using Eq. 11, the transition voltage, $V_t$, can be extracted from the bias where a given trace crosses the NDC=2 line (dashed horizontal line in Fig. 7C). Such NDC derived $V_t$ values are shown in the right-most column of Table 1. It shows that the two methods ($V_t$ by TVS or NDC=2) are practically identical except for minor noise fluctuations.

However, in contrast to the original TVS approach, different curves can be compared, even if they do not reach the specific value of NDC=2. If for technical reasons the available bias range is limited, then a different arbitrary value can be chosen for comparison (e.g., NDC=1.5), and translated into equivalent $V_0$.[365,369] In addition, the NDC presentation clarifies that the quasi-Ohmic curve of Br-Styrene has NDC<1.2 at any $|V|<1$ V, confirming the Ohmic nature of this junction. Therefore, the NDC presentation combines the benefits of both $V_t$ and parabolic *G-V*: it provides both meaningful curve shape and unambiguous extraction of a scaling parameter. Exact relations between NDC and specific transport physics as well as relation to other electronic-spectroscopies like IETS and opening of additional transport channels (e.g., switching or rectification ratio discussed next) are yet to be explored.

Accumulating data show that MolEl junctions with conjugated molecules have a scaling voltage of 0.5 V or less,[113] while saturated alkyls have a higher scaling voltage, on the order of 1 V or higher.[347,367] As noted above, the reciprocal length dependence of $V_0$ is still disputed.[347] In comparison to those values, the extracted $V_0$ values for Si-alkyl/Hg (Table 1) are very low. This together with the fact that $V_0$ depends on the doping type of the Si, suggest that the shape of the *I-V* curves in these Hg/alkyl-Si junctions is dictated by the availability of charge carriers from the Si (despite its nearly degenerate doping), combined with the transmission function of the molecules. This might explain why the single Lorentzian failed (Fig. 8B) to describe this type of junctions.

To summarize, the Taylor expansion justifies the notion that tunneling I-V curves behave like scaled relations. The characteristic, scaling voltage can be extracted by various methods that differ basically in the choice of how close to 0 V this quantity is extracted. We argue that the scaling voltage, $V_0$, should be added to the other empirical observables of transport by tunneling ($G_{eq}$, $\beta$ – the length decay, and $E_a$ – the temperature dependence). The scaling voltage has no single interpretation; the same holds for the activation energy, $E_a$, which can have different meaning depending on our *interpretation* of the transport process (e.g., hopping or level broadening, see section 5.2).

# 6. Structure-Function relations in Molecular junctions

## 6.1. Rectification and switching

For a molecular junction to be functional electrically, it should be 'responsive', i.e., change its electrical conductance as a response to some input signal. The main control signal of traditional electronics is the applied bias, though memory elements are often driven by magnetic or optical stimuli. In principle, molecules can be sensitive to all these stimuli, and others such as pH or chemical recognition in general, and there are a variety of experimental demonstrations of such effects.[3,4] Conformational isomerization is a powerful switch in biological molecules and often observed in solution, but attempts to observe it in solid junctions gave rather disappointing results and were and poorly reproducible. This was attributed to either geometrical constraints in single molecule junctions[5] or steric limitations due to dense packing of neighboring molecules in monolayer junctions.[370] We find that hybrid molecular / semiconductor junctions are often more reliable in transducing chemical recognition[371-373] than pure molecular devices.

Shifting the molecular energy levels with respect to those of the electrodes is possible using a 'gate-electrode' - a third electrode in close proximity to the junction, but not contacting it. Gating is very appealing to gain insight into molecular energy levels, as is often done for single molecules in either electro-migrated contacts[4,289] (via an underlying gate insulator) or in liquid STM (by electro-chemical gating).[2,287,288] While these configurations are often named 'transistors' they do not provide a power gain, as is known for the gated electrical device, the transistor.[1] Introducing a gating field in NmJ is highly non-trivial in view of the unfavorable aspect ratio of monolayer (large area, small width) junctions.[374-376] Here we focus on direct characteristics of large-area molecular junctions, and therefore we shall limit the discussion to non-linearity in two-terminal devices.

The standard two-terminal electronic device, which shows non-linearity with applied bias, is a diode: it has a high conductance when bias of one polarity (e.g., +1 V) is applied and a very low conductance with a bias of the opposite polarity (e.g., -1 V). This is called 'rectification', namely the current under the reverse ('wrong') sign of bias is suppressed compared that for the forward bias polarity. In addition, a simple two-terminal device can show a variety of other non-linear effects, such as negative differential resistance (NDR),[377] hysteresis and switching.[378] A diode-like rectification is actually a specific case of switching which occurs near 0 V, while, in theory, molecular electronics may offer a way towards multiple switching events (conductance onsets) at various bias levels. A bias-symmetric switch is for example a sharp onset between low and high conductance at an arbitrary voltage, $V_{onset}$, and its mirror onset at $-V_{onset}$, to yield three conductance regimes: high / low / high for large negative, small bias and large positive bias, respectively. For example a porphyrin island within an insulating alkyl-thiol matrix, shows a conductance onset around ±0.1 V, where the current above this voltage is up to 1000 fold larger than near 0 V.[195] Large area, carbon / conjugated molecules / metal junctions also show sharp conductance onsets that can be used for music amplification.[31] Here we restrict ourselves to the simplest mode of non-linearity, rectification, expressed via a simple empirical ratio, known as 'rectification ratio', $RR$:[332,379]

$$RR(V_i) = |I(+V_i)/I(-V_i)| \qquad (13)$$

where $V_i > 0$ by definition and the sign is normally chosen to yield RR>1. Within the sign convention of Eq. 13, Aviram and Ratner originally predicted RR>1 when the acceptor is grounded (i.e., V > 0 on the donor side is the 'ON' state);[11] however experimental molecular donor/acceptor junctions consistently show an opposite rectification direction[1] (i.e., RR > 1 for V < 0 on the donor side, or the 'HOMO'-side of the molecule is shifted upward in the energy diagram[380]). Such an inversion was recently explained by charge equilibration between the molecular donor / acceptor moieties and the electrodes.[362] From the functional perspective, the rectification ratio alone is not sufficient to describe switching, and other critical features are how sharp is its onset, how wide and stable are the low/high conductance windows, what is the switching speed, retention time and fidelity.

Rectification in molecular devices is thoroughly reviewed recently by Metzger,[1] who compiled a few dozen reported molecular rectifiers. Most of them have $RR$< 10, and several are between 30 and 100.[1] Such low rectification is probably better termed asymmetry (of the I–V curve). A very few molecular junctions have reached $RR \geq 1000$, using a donor-acceptor,[381] ferrocene redox center,[382], or with a C60 monolayer.[383]

There are a few different possible mechanisms that explain rectification, which involves either specific donor / acceptor levels in the molecule or just the built-in field within the film (see Ref. 1 for a detailed explanation). Notably, all examples of RR ≥ 100 were demonstrated for monolayers (NmJ) but not with single molecule junctions (1mJ). A possible reason for this NmJ / 1mJ difference is that monolayers, in contrast to single molecules produce huge internal electric fields (1-100 MV/cm) due to the collective dipole effect.[51,54,327,384] The importance of the built-in field in inducing rectification was recently demonstrated for a 1mJ where a *symmetric* molecule yielded $RR \sim 100$.[385] The asymmetric field is induced by the different geometry of a sharp STM tip vs. the flat relatively large area bottom substrate, that induced polarization of a surrounding electrolyte.[385] This work points to the critical role of electric field lines in dictating electrical asymmetry across interfaces.[386] It is in line with the large difference in field lines observed by DFT computations for junctions made of identical molecules, either in a single-molecule configuration or an array configuration.[54]

It suggests that large electric fields built-in to the junction are one promising route to achieve transport rectification. The mechanism to achieve this strong built-in field could be intrinsic molecular dipole (only for arrays)[54,387] or tip-geometry plus electrolyte environment.[385] Internal fields are also critical for the donor-acceptor based rectification mechanism.

Nevertheless, $RR$ values of metal/molecule/metal junctions are still far from those of semiconductor-based junctions. If a moderately doped semiconductor is used as one contact in MolEl, the rectification is controlled by the barrier within the semiconductor, as discussed elsewhere.[46,85] However, also if a highly doped semiconductor is used in MolEl, some rectification can originate from the semiconducting nature of the contact. As shown in Fig. 7, a junction made of Hg/C16-Si, where the Si is highly-doped p-type leads to highly non-symmetric transport, with a maximal $RR$ of 8.5 at V=±0.75, compared to only 1.4 for exactly the same junction only made with highly-doped n-type Si. A similar highly-doped Si, which is often considered as effectively metallic, was recently reported to yield $RR$ of $10^6 - 10^7$, when the molecular layer contained an active redox moiety (porphyrin), with Hg as a top contact.[388] These huge $RR$ values for porphyrin-based monolayers must originate in a unique synergism between the molecules and the highly-doped Si. Indeed, the giant $RR$ of $10^6 - 10^7$ was observed only if the porphyrin was tethered to the Si via 11 carbon long alkyl chains, while a 6 carbon tether yielded only $RR \leq 100$.[388] Overall, it is an inspiring example for the synergetic prospects of hybrid monolayer/Si interfaces.

In the case of a passive alkyl monolayer, we explained the asymmetry in analogy to a metal-insulator-semiconductor 'Esaki diode' (related to, but different from the more familiar p-i-n Esaki diode).[389] In this mechanism, rectification (and ideally even negative differential resistance) occurs when there is a mismatch in the type of dominant carriers of the insulator (molecules here) and the semiconductor. UPS reveals that the frontier level of alkyl monolayers is their LUMO,[348,390] namely electrons dominate the tunneling through the Si-bound alkyl. Therefore, for an alkyl monolayer adsorbed onto highly-doped n-Si the dominant carriers are the same for the semiconductor and the insulator and roughly symmetric I-V curves are obtained. In contrast, the junction that is identical except for that it uses p-Si, has opposite types of carriers for the molecules and the semiconductor and therefore asymmetric transport.[391] This asymmetry is clear using the normalized differential conductance presentation (NDC, Fig. 7C) which even yield NDC<1 (though not negative).

While a semiconducting electrode enhances the sensitivity to internal potential profiles, the ability to induce an asymmetric potential profile across nm-long molecules is critical for achieving asymmetric transport.[392,393] The choice of contacts and binding groups is the main handle to manipulate the internal potential profiles, as discussed next.

## 6.2. Contacts and Energy Alignment

While an original vision of molecular electronics was one of exploiting molecules as naturally isolated quantized systems, the recent view is one of molecule + contacts as the actual chemical entity.[5] The degree of interaction between the molecule and the contacts is largely controlled by the binding / terminal groups, connecting the core of the molecule to the contacts. However, a strong chemical bond, does not necessarily imply efficient electronic coupling; for example, if bonding causes strong electron-localization, as e.g., in ionic bonds, it hampers the interface transmission. Proximity alone can induce energy-level hybridization and considerable charge rearrangement even for a mechanical contact, that would normally be considered as at most physisorption (i.e., no covalent, ionic or metallic chemical bond is formed).[8,47,394-398] At the same time, chemical binding ('chemisorption') leads to largely different energy coupling (high $\Gamma$ values, see Fig. 6), depending on the type of bond and the details of the binding groups.[30,48,54,211] Therefore, there is no simple relation between the binding strength (the physi- / chemi-sorption distinction) and the levels hybridization and coupling across the interface.

The key features describing the transport (see Eqs. 2, 8 and Fig. 6A) are the energy alignment of the frontier molecular levels with respect to the contact's Fermi level, $\varepsilon_0$ ($\sim V_0$) and the strength of the coupling of these levels to the density of states of the contacts, $\Gamma$ ($\sim G_{eq}$). Both characteristics are strongly affected by the chemical interactions between the molecule and the substrate by equilibration of the electro-chemical potential of the electrons and overlap of the frontier orbitals, as dictated by stereochemistry and symmetry. Both $\varepsilon_0$ and $\Gamma$ are closely related to surface adsorption processes as argued in Ref. 8. We refrain from using the term 'binding' so as to encompass the wide variety of interactions that can exist between molecules and the contacts (see Refs. 1,5 for detailed list; cf. also section 3.1). As the 'core' of the molecule or its central part is often homogenous (e.g., alkyl chain or fully conjugated, non-substituted wire), it is, therefore, normally not expected to be polarized, unless specifically engineered. At the same time, there are many examples of molecules with identical core moiety (e.g., phenyl or alkyl), for which variation of their end groups change their energetics[48,211,306,393,399,400] and transport.[5,30,42,346,355,401-403]

It takes two to tango and the molecule / contact interaction is dictated by both the end group and the nature of the contact.[5] The strength of the electrical coupling, $\Gamma$, is affected by the extent of overlap between the atomic orbitals of the contact and the molecular frontier orbitals,[5] which depends on their geometry relative to each other and, possibly the orbital symmetry and spin. The number of conductance channels relates to the number of valence electrons for a given metal.[404] For example, both pyridine and amine bind to Au via the lone-pair electrons of the nitrogen. While the binding strength of pyridine to Au is larger than that of amine, it yields a lower net conductance because the nitrogen lone-pair in pyridine is orthogonal to the π-system.[30] Coupling is specific for the different orbitals of a molecule, and efficient coupling can even favour transport from deeper orbitals, rather than from the one nearest to the Fermi level, if the latter has far poorer coupling than the former.[405] Still, the coupling term, Γ is not only sensitive to stereo-chemistry but also to the energy difference. Here we focus on devising guidelines for the energy alignment between the molecular energy levels and the electrode's Fermi level.

Compared to inorganic solids, molecules are characterized by spatially localized orbitals and low dielectric constant. As a result, molecules can tolerate huge electrical potential differences over atomic distances (cf. 100's nm in inorganic semiconductors) and can, therefore, be viewed as "atomic-scale hetero-structured devices."[393] This characteristic is expressed in large area junctions (NmJ), where the individual dipoles are effectively screened by neighboring dipoles, leading to extremely sharp potential steps along each molecule,[51,400] Such potential steps are constrained to just one or two atoms immediately next to the solid / molecule interfaces (i.e., the end group, regardless of whether a chemical bond exists or not),[400] a step that determines to a large extent the relative alignment of the molecular energy levels with the Fermi level of the electrodes.[43,54,113,386,387,400,406] The lack of collective electric-field screening in 1mJ implies that a different potential profile is developed across the same molecule when connected within 1mJ or NmJ, with a milder potential gradients across 1mJ compared to NmJ.[54] Yet build-up of internal potential profiles is critical also for 1mJ.[362,393]

It is now clear that varying the metal work function has a minor effect on the energy barrier height ($\varepsilon_0$ in Fig. 6)[43,53,113,400] though the coupling term, $\Gamma$, depends exponentially on the work function.[42,43,111,113] The first observation (metal-independent $\varepsilon_0$) implies that the molecular energy levels maintain a roughly constant alignment with the metal's Fermi level, regardless of how deep is the metal Fermi level relative to vacuum. Thus, the reference energy is the system's Fermi level rather than the vacuum level. In other words, upon adsorption the electro-chemical potential of the molecule becomes aligned with the Fermi level of the contacts. Therefore, the difference between the frontier orbital and the electrochemical potential of the *molecule* (rather than the absolute ionization potential) dictates the barrier, almost independently of the contacts.[43,113] The concept of 'rigid' electrochemical potential is in line with the observation that the work-function of monolayer-modified surfaces (saturated[399,407] or conjugated[43,113]) is roughly the same even though the work-functions of the clean surfaces vary by ca. 1 eV. Namely, the molecularly-induced dipole is the result of some (electronic) thermodynamic equilibrium between the molecule and the substrate and not an intrinsic property of the molecule. Thus although molecules are extremely small objects, there is a clear partition between the 'end region' and the core. This invoked suggestions that transport is better described by a series of three[113,149] or multiple barriers,[408] rather than a single one.

The second observed effect of the metal's work function is on the coupling, $\Gamma$, with the molecule (or $G_{eq}$ for $L \rightarrow 0$), which is measured experimentally as contact conductance, $G_C$:

$$G = G_C \cdot exp(-\beta L) \qquad (14)$$

where $\beta$ is the 'exponential decay parameter', and both $\beta$ and $G_C$ can have different physical descriptions.

While the work-function is not considered in standard transport models, Frisbie and co-workers have repeatedly observed an exponential increase in $G_C$ for junctions of given molecules with an increase in work function of the contacting metal (mostly for thiol binding but also for physisorbed contacts).[42,43,111,113] In principle, better energy matching between the end group and the electrode could enhance the coupling, however, the observation is that efficient coupling occurs for large induced dipole[43,113] (i.e., apparently an energy mismatch), which is not trivial. The correlation was specifically with the work function of the electrodes and not with their elemental electronegativity, suggesting that the energy alignment involves the complete system and not only the binding atoms (e.g., metal – sulfur). The DOS of the clean metal could also not explain this trend, suggesting that some induced hybrid resonances[8] might be involved here.

The work function logic suggests that chemical alteration of end groups should also strongly alter the transport probability. Yet, Whitesides and coworkers have systematically varied the binding[409] or terminal group,[410,411] using the Ag-monolayer /InGa configuration, and found no significant change in transport probability. While both examples are for NmJ, it cannot be ruled out that the measurement configurations (presumably very clean conductive tip AFM vs. InGa with a thin oxide buffer) play some role.[57] Interestingly, the weak sensitivity to end groups in NmJs is in contrast to the great sensitivity of single molecule junctions (1mJ) to the end groups.[5,30,346,403] The reason is probably the efficient screening of electric fields in NmJ's[51,53,400] which is absent in 1mJs.[54] It is quite probable that the difference in the penetration of the electric field also influences the level hybridization[54] ($\Gamma$ in Fig. 6) and leads to differences in transport characteristics between 1mJ and NmJ.

## 6.3. Length dependence and other considerations

Quantifying molecular transport by its length sensitivity is a very common practice. For transport by non-resonant tunneling (including a variety of specific mechanisms) the conductance decays exponentially with the tunneling distance, $L$ (Eqs. 3 and 14 above).[245]   Although eq. 14 is widely used, there are some subtle issues that have to be noticed. The first one is the assumed identity between tunneling distance and molecular length. This identity is not obvious especially when the end-sites are not well-defined (e.g., for proteins) or for physisorbed molecules,[136] and more so for most 1mJ experiments where the absolute distance between the contacts is externally imposed and is mostly difficult to calibrate. In addition, Eq. 14 implies that extracting $\beta$ by comparing a series of length-varying homologs, alters solely the length while all other characteristics are unaltered. This would be inaccurate in some cases. For example, increasing the length of a fully conjugated system decreases its HOMO-LUMO gap and by that alters $\varepsilon_0$,[43,113,412] while varying the length of alkyl phosphonic acids alter their tilt and packing.[86]

Bias is another complication. Although the voltage is not specified in Eq. 14, all its terms except of length (i.e., $G$, $G_C$, $\beta$) are bias-dependent; this complication is conveniently bypassed by comparing the conductance at 0 V (i.e., $G_{eq}$ of Eqs. 2, 3). Still, extracting $\beta$ using current at applied voltage ≠ 0 is often found in the literature, a practice that introduces some ambiguity. Actually, there are contradicting examples for the voltage (V) effect on $\beta$ values. While some papers report a clear dependence[197,253,324,413] of $\beta^2 \propto -|V|$ others found a voltage-independent $\beta$.[111,113,245] It is our understanding that a voltage-sensitive $\beta$ occurs in cases of significant potential drop on the molecule (i.e., trapezoidal potential profile) while constant $\beta$ is characteristic of potential drop restricted to the contacts (a step profile).[347]

Returning to the $G_{eq}$ / $\varepsilon_0$ description (i.e., the generic partition discussed in section 5.3), the common assumption is that the distance attenuation is contained within the equilibrium con-ductance term, $G_{eq}$ (or Γ in transmission function terminology of Fig. 6). Most tunneling models predict some relation between $\beta$ and the injection barrier (e.g., $\beta \sim \sqrt{\varepsilon_0}$ in Simmons' model,[253,324,333] or $\beta \sim \ln(\varepsilon_0)$ for super-exchange.[40,111] Therefore, $G_{eq}$ and $\varepsilon_0$ are possibly correlated, an assumption that can be tested experimentally,[272] to gain additional verification how suitable is a given transport model. Tunneling is expected to be more efficient along conjugated molecules, and indeed $\beta$ values for conjugated molecules (ca. 0.2-0.3 $Å^{-1}$)[19] are significantly lower[245] than those of alkyl chains (ca. 0.8 $Å^{-1}$).[19,28] The prevailing similarity in $\beta$ values for alkyls[5,28] suggests that basically, the injection barrier, $\varepsilon_0$ is very similar regardless of vastly different binding groups and contacts.[53,111,409-411]

Overall, we still miss guiding rules of the thumb which molecular features are more expressed in charge transport and which less. The strategy that now appears most effective in tuning transport across NmJ is a combination of saturated and conjugated units, such as embedded ferrocene,[380] or phenyls,[408] within alkyl segments of varying length. There is accumulating evidence that as few as 1-2 methylene units are sufficient to buffer between 'active' π systems and the electrodes[48,414,415] or between two molecular conjugated moieties.[11,362] Such saturated buffers are critical for breaking the 'equi-potential' state, toward realization of pronounced transport asymmetry or electronic functionality. Addition of hetero-atoms and specific donor / acceptor groups within the molecule could induce pronounced asymmetry in NmJ[380,382,386,387] but apparently much less so in 1mJ.[415,416]

## 7. Summary

We have reviewed the current understanding of electrical charge transport across molecular monolayers, a branch of molecular electronics (MolEl), which studies charge transport across numerous molecules in parallel (NmJ). We showed that, compared to a single molecules (1mJ) junction, a NmJ is closer to chemical equilibrium and shows pronounced collective effects, leading to stronger built-in electric fields and, therefore, more pronounced non-linear response to applied voltage. Construction of NmJs must consider both the monolayer and the contacts, with prime importance to diffusion barriers, roughness, defect healing, and molecular flexibility. NmJs are also of interest for testing artificial bottom-up self-assembly toward mimicking of bio-systems.

Charge transport across short molecular junctions is tunneling-dominated. This leads to an apparent paradox where formal theory requires detailed atomistic description, while in practice the resulting I-V traces have fairly similar shape, regardless of molecular details. We showed that derivative approaches, such as plotting the differential conductance (dI/dV) or resistance (V/I) against the applied bias can provide some fingerprints for specific tunneling mechanisms, and to identify practical ranges in terms of applied bias or temperature where a given molecular junction can be described by a specific tunneling version (e.g., coherent vs. sequential tunneling).

The inherent similar shape of tunneling-dominated I-V traces can be understood as a scaled relation, where the current magnitude is scaled by the zero-voltage conductance ($G_{eq}$) and the scaling bias, $V_0$, measures how non-linear (non-Ohmic) is the I-V trace. In this sense, $V_0$ is similar to an activation energy (effect of temperature) or decay length (β). All are empirical observables, which are not limited to, or inform about a definite transport mechanism. We discussed several types of analyses for extracting $V_0$ and recommend normalized differential conductance (NDC) as a robust, quantitative way to extract $V_0$ and as qualitative expression of voltage effect on transport.

Finally, we briefly covered how to introduce non-linearity or electrical functionality into molecular junctions. While the binding / terminal groups are often considered in terms of coupling strength to the electrodes, we emphasize their role in dictating the net energy alignment and build-up of an internal electric field. This internal field may break the molecular 'equi-potential' state and by that induce significant bias-asymmetry toward functional MolEl.

## AUTHOR INFORMATION

### Corresponding Author

*E-mail: ayelet.vilan@weizmann.ac.il

### Notes

The authors declare no competing financial interest.

### Biographies

**Ayelet Vilan** received her Ph.D. in the Materials and Interfaces department of the Weizmann Institute of Science, studying dipolar monolayers at the Au/GaAs surface. She did post-doctoral studies at Philips Research and Ben-Gurion University of the Negev, after which she returned to the Weizmann Institute in 2005, adding a specialization in surface science. In 2013 she spent a year at Texas A&M. Her research interests focus on molecular electronics, until recently primarily on large-area molecular junctions, especially those on semiconductor electrodes and more recently also include single-molecule break junctions.

**Dinesh Aswal** is the Director of the CSIR- National Physical Laboratory, New Delhi. He served as the Secretary of the Indian Atomic Energy Education Society, Mumbai and as Head, Thin Films Devices Section, Technical Physics Division, Bhabha Atomic Research Center (BARC), Mumbai, which he joined at 1986 after completing an M.Sc. in Physics (Gold medalist) from Garhwal University in 1985. He obtained a Ph.D. in Physics from Mumbai University for work on "Thin films of high temperature superconductors", and carried out post doctoral research at the Research Institute of Electronics, Hamamatsu, Japan. His current area of research interests includes physics of organic monolayers, conducting polymer films for flexible electronics, thermoelectric power generators and gas sensors & electronic nose. He has edited three books, contributed 18 book chapters and published over 200 peer reviewed journal papers and has received numerous awards and prizes.

**David Cahen** studied chemistry & physics at the Hebrew Univ. of Jerusalem (HUJ), Materials Research and Phys. Chem. at Northwestern Univ, and Biophysics of photosynthesis (post-doc) at HUJ and the Weizmann Institute of Science, WIS. After joining the WIS faculty he focused on alternative sustainable energy resources, in particular on various types of solar cells. Today his work in this area focuses on the materials and device chemistry and physics of high-voltage (mainly halide perovskite-based) cells. In parallel he researches hybrid molecular/non-molecular systems, nowadays focusing on peptide & protein bio-optoelectronics and implications for electron transport across biomolecules. A fellow of the AVS and MRS, he heads WIS' Alternative, sustainable energy research initiative.

potential of the electron ($\Delta\mu e$) between donor and acceptor. Electrochemistry across monolayers is driven by the difference in the electrons' electrochemical potential ($\Delta\eta_e = \Delta\mu_e + \Delta\phi$). See also Refs. 17, 19.

Waals forces) of the same kind as those responsible for the imperfection of real gases and the condensation of vapours, and which do not involve a significant change in the electronic orbital patterns of the species involved. The problem of distinguishing between chemisorption and physisorption is basically the same as that of distinguishing between chemical and physical interaction in general. No absolutely sharp distinction can be made and intermediate cases exist, for example, adsorption involving strong hydrogen bonds or weak charge transfer." Taken from:

http://old.iupac.org/reports/2001/colloid_2001/manual_of_s_and_t/node16.html